\documentclass[journal]{new-aiaa}
 \usepackage[utf8]{inputenc}

\usepackage{graphicx}
\usepackage{amsmath}
\usepackage{physics}
\usepackage[version=4]{mhchem}
\usepackage{siunitx}
\usepackage{longtable,tabularx}
\usepackage{hhline}
\usepackage{fancyhdr}
\fancypagestyle{firstpage}{%
  \lhead{\textit{This preprint has been submitted to the AIAA Journal, and therefore third-party reuse of any published material requires permission from AIAA.}}}

\usepackage{wrapfig}
\usepackage{subfigure}
\usepackage{subfigmat}
\usepackage{xcolor}

\newcommand*{\rttensor}[1]{\bar{\bar{#1}}}

 \title{Pressure Bifurcation Phenomenon on Supersonic Blowing Trailing Edges}

 \author{
  Alejandro Martinez-Cava\footnote{Early Stage Researcher. Correspondence to alejandro.martinezcava@upm.es}}
 \author{Yinzhu Wang\footnote{Early Stage Researcher. Correspondence to y.wang@upm.es}}
 \author{Javier de Vicente\footnote{Associate Professor of Applied Mathematics. Correspondence to fj.devicente@upm.es}}
 \author{Eusebio Valero\footnote{Professor of Applied Mathematics. Correspondence to eusebio.valero@upm.es}}
 \affil{School of Aeronautical and Space Engineering, Universidad Polit\'ecnica de Madrid. \\
 Plaza Cardenal Cisneros 3, E-28040, Madrid, Spain}

\begin{document}

\maketitle
\thispagestyle{firstpage}
\begin{abstract}

Turbine blades operating in transonic-supersonic regime develop a complex shock wave system at the trailing edge, a phenomenon that leads to unfavorable pressure perturbations downstream and can interact with other turbine stages. Understanding the fluid behavior of the area adjacent to the trailing edge is essential in order to determine the parameters that have influence on these pressure fluctuations.  Colder flow, bled from the high-pressure compressor, is often purged at the trailing edge to cool the thin blade edges, affecting the flow behavior and modulating the intensity and angle of the shock waves system. However, this purge flow can sometimes generate non-symmetrical configurations due to a pressure difference that is provoked by the injected flow. In this work, a combination of RANS simulations and global stability analysis is employed to explain the physical reasons of this flow bifurcation. Analyzing the features that naturally appear in the flow and become dominant for some value of the parameters involved in the problem, an anti-symmetrical global mode, related to the sudden geometrical expansion of the trailing edge slot, is identified as the main mechanism that forces the changes in the flow topology.

\end{abstract}


\section*{Nomenclature}

{\renewcommand\arraystretch{1.0}
\noindent
\begin{longtable*}{@{}l @{\quad=\quad} l@{}}
$\beta$  & spanwise wavenumber \\
$\rho_f$  & density measured at the \textit{base region} \\
$\rho_{purge}$  & density measured at the cavity plenum \\
$\Omega$  & control volume \\
$\Omega_i$  & mesh subdomain \\
$\omega$  & complex eigenvalue \\
$\omega_i$  & eigenvalue imaginary part (pulsation) \\
$\omega_r$  & eigenvalue real part (amplification rate) \\
$\mathbf{A}$  & linearized Jacobian matrix \\
$\mathbf{B}$  & volume matrix \\
$d$  & trailing edge thickness \\
$f$  & frequency \\
$\rttensor{F}$  & flux tensor \\
$k$  & turbulent kinetic energy \\
$N$  & number of grid points \\
$\mathcal{N}_f$  & number of faces on subdomain \\
$N_v$  & number of fluid variables \\
$\mathbf{n}$  & normal direction to the body surface \\
$\mathit{omega}$  & specific dissipation rate of $k$ \\
$\mathbf{P}$  & projection operator matrix \\
$P_b$  & pressure measured at the \textit{base region} \\
$P^*_b$  & pressure measured at the \textit{base region} when no purge flow is applied \\
$\mathbf{Q}$  & \textit{base flow} vector field \\
$\mathbf{q}$  & conservative variables vector \\
$\mathbf{q_i}$  & discretized vector state solution in subdomain $\Omega_i$ \\
$\mathbf{\widetilde{q}}$  & vector of small perturbations \\
$\mathbf{\widehat{q}}$  & vector of complex disturbances \\
$\mathbf{R_i}$  & residual vector in subdomain $\Omega_i$ \\
$res^n_p$  & global density residual  \\
$St$ & Strouhal number \\
$T$ & temperature \\
$t$  & time \\
$u,v,w$  & velocities in x-,y- and z-directions respectively \\
$\mathbf{x}$  & vector of coordinate directions (x,y,z) \\
$y^+$  & dimensionless wall-normal distance \\
$z$  & spanwise direction
\end{longtable*}}


\section{Introduction}

\lettrine[nindent=0pt]{A}{IR} transport has experienced a very large growth in last decades, with a significant socio-economic relevance for integration and development at regional and international level. From a technological point of view, the aerospace industry is constantly at the cutting edge of scientific development. Continuous improvements in aerodynamics, material sciences, manufacturing processes, avionics, control and navigation systems make aircraft more efficient and safer every day. In particular, aircraft engines tend towards more compact architectures, the structural safety limits are tighter and aero-structural couplings may lead to low and high cycle fatigue issues. Such aero-structural interactions are significantly detrimental when the internal flow velocities approach the speed of sound.

Turbomachinery airfoils can operate in transonic or supersonic flow regime, experiencing complex shock wave systems at the trailing edge that can lead to unfavorable pressure perturbations affecting the life-span of the turbine components. Part of the aero-structural coupling has its origin on the flow fluctuations that take place at the area adjacent to the trailing edge, known as the base region. The evaluation of the parameters driving the compression and expansion waves, shear layers and vortex shedding should then allow us to mitigate their unsteady couplings with the turbine blades.

\begin{wrapfigure}[16]{r}{0.4\textwidth}
  \centering
    \includegraphics[width=0.38\textwidth]{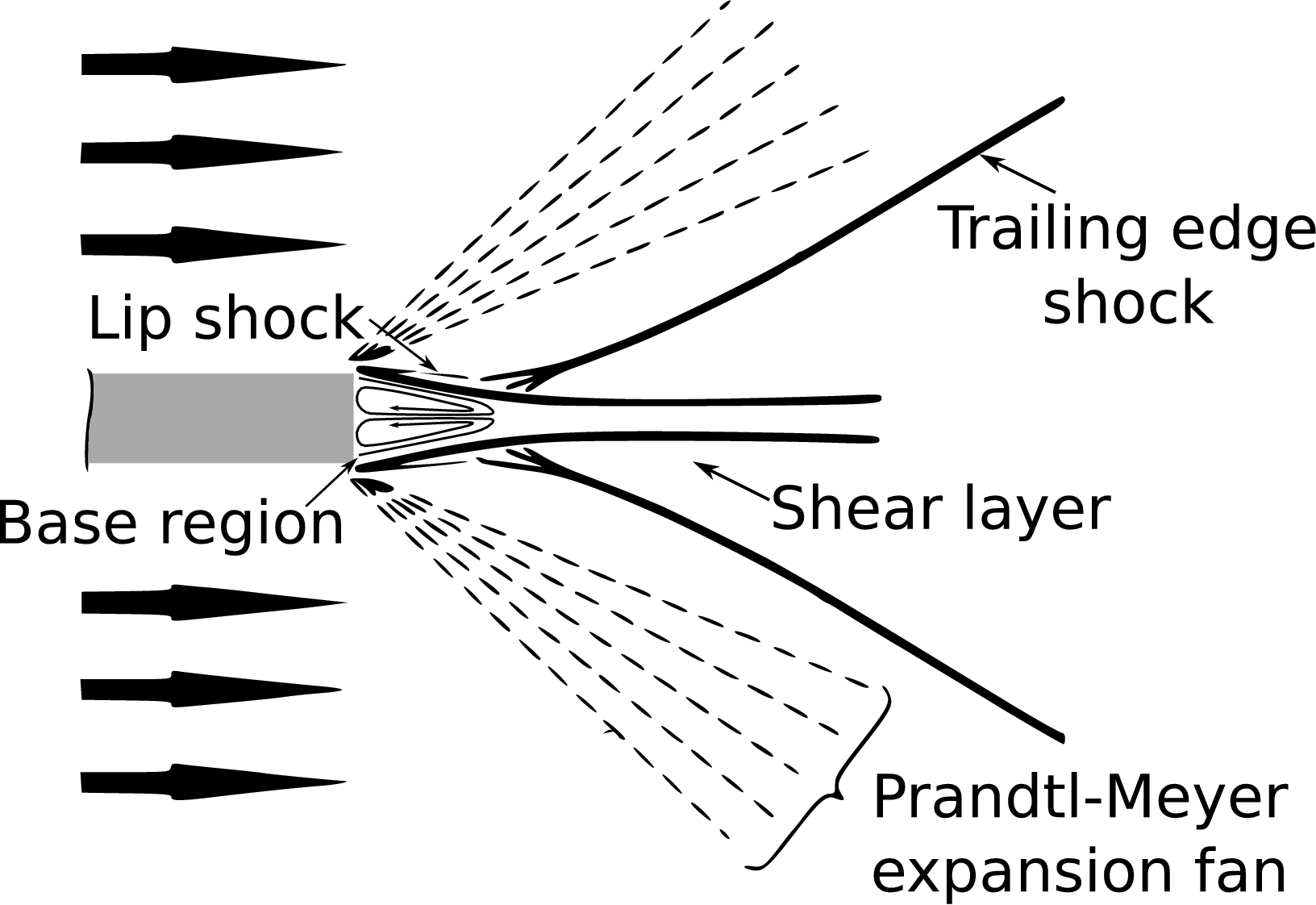}
  \caption{Schematic representation of supersonic flow on a blunt trailing edge. Image modified from Saracoglu et al.\cite{Saracoglu2013}}
  \label{f:intro:scheme}
\end{wrapfigure}
Figure \ref{f:intro:scheme} summarizes the flow topology of a supersonic trailing edge, often approximated by a simplified model based on the supersonic flow over a flat plate end. The upstream boundary layer remains attached at the trailing edge corner, and separates immediately downstream developing a shear layer while, simultaneously, the main flow accelerates through a Prandtl-Meyer expansion fan. The upper and lower shear layers propagate downstream and eventually merge at the wake on a point of confluence, creating a dead air zone limited by both shear layers and the blunt trailing edge, namely the base region. This zone is characterized by a low momentum and constant pressure, having a strong influence on several features of the surrounding flow field, such as the vortex oscillation intensity or  shock waves position and angle. At the confluence point, the flow changes its direction notably, being forced to compress through a system of strong trailing edge shocks. The degree of compression, hence the strength of the shock waves, highly depends on the base region properties. Furthermore, a weak compression wave, called separation or lip shock, is formed at the point of flow detachment in order to adapt the pressure gradient between the shearing and the downstream flow. The base region has been an object of study for long time \cite{Nash1963,Nash1967,Nash1966,Hama1968}, and its influence and characteristic topology on turbine cascades\cite{Sieverding1979,Denton1990} or the supersonic turbulent wake of axisymmetric bodies\cite{Herrin1994,Sandberg2012,Sandberg2006} repetidly revisited.
\begin{figure}
	\centering
	\includegraphics[height=4cm]{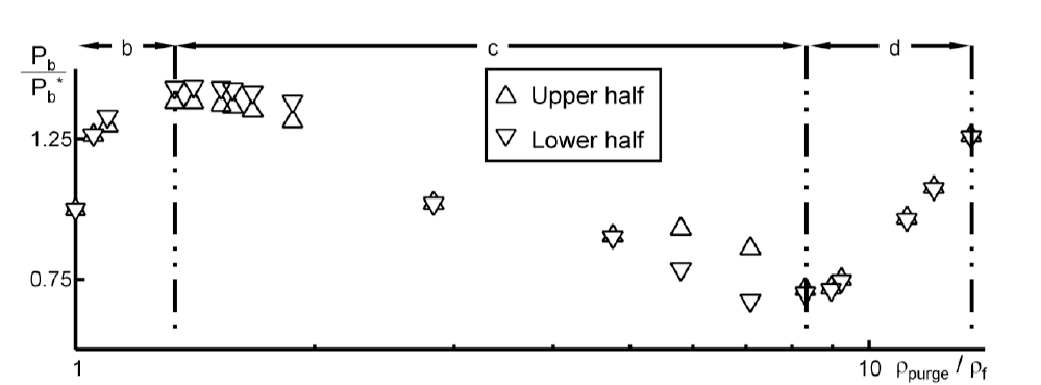}
	\caption{Base pressure correlation at the trailing edge as a function of the density ratio of the bleeding rate. Image reprinted from Saracoglu et al.\cite{Saracoglu2013}.}
	\label{f:ref_guillermo}
\end{figure}

The thin trailing edge of turbine blades cannot be protected using internal cooling passages, so cooler flow bled from the high pressure compressor is purged through slots or orifices to avoid the melting of the turbine components. As it has been shown by several authors\cite{Saracoglu2012,Raffel1998}, this cooling could be exploited to modulate the base region properties. Different studies\cite{Kost1985,Bohn1995,Wang2013} have been devoted to examine the effects of trailing edge bleeding on the base region. Of particular importance, Saracoglu et al.\cite{Saracoglu2013} investigated the flow topology at the blunt trailing edge as a function of the cooling flow intensity, showing the dependence of the base pressure with the blowing magnitude. A relevant result of the study was the presence of non-symmetric configurations at the base flow (shock intensity and angle of the shock wave) for a specific range of blowing rates (Figure \ref{f:ref_guillermo}). Two different bifurcations from symmetric to non-symmetric configurations were observed, the first one arising when the base pressure is close to reach its maximum value, and a second one appearing for higher blowing ratios, close to a local minimum value for the base pressure.

The objective of the present research is to combine RANS simulations and linear stability theory to explain the mechanisms that generate the described non-symmetric configuration. Only the first pressure bifurcation is analyzed in detail, focusing on the area where the purge flow provokes an important increment on the base pressure, reducing the intensity of the shock waves and the base pressure losses. A thorough analysis of the flow topology of a supersonic blowing trailing edge is carried, focusing on the recirculation areas of the base region, showing how those areas interact generating changes on the injected flow and therefore in the complex wave system. Those changes are then also identified using global stability analysis. Stability analysis studies the growth or decay of perturbations superimposed upon a steady base flow, and has proven to be effective in the analysis of incompressible and compressible flows, either in laminar or turbulent configurations. The analysis aims to identify which features can naturally appear in the flow and become dominant for some particular value of the parameters. Depending on the nature of the phenomenon, the analysis can be performed over a steady \textit{base flow} solution\cite{Iorio2014,Iorio2015,Sartor2015}, over an averaged \textit{mean flow}\cite{Barkley2006,Oberleithner2014,Sipp2007}, or using time-stepping or DMD techniques for the study of transient phenomena\cite{Grilli2012,Lashgari2014}. On this work, steady \textit{base flows} were considered to characterize the stability behaviour of the system. The results of this study shed some light on the pressure bifurcation detected by Saracoglu et al.\cite{Saracoglu2013}, identifying the physical global mode responsible of the non-symmetrical conditions as a function of the cooling flow purge intensity. We anticipate that results of the stability analysis match the bifurcation predicted by RANS simulations, adding extra information about the underlying physical mechanisms. The unstable configuration seems to be related to the sudden geometrical expansion of the flow at the end of the cooling slot. The associated instability forces the purge flow direction to deflect and impact one of the shear layers, generating an additional shock wave system. Such changes on the vicinity of the trailing edge would have a great effect on the loads affecting adjacent blades and turbine stages. The identified global mode changes its shape with the blowing rate intensity, as does the flow topology of the base region, giving the stability analysis valuable information about where the perturbations are located.

The rest of the paper is organized as follows. Section II overviews the mathematical formulation of the stability analysis and implementation details, opening the door to future applications in flow control. Section III summarizes the numerical procedures to obtain the base flow and stability solutions. In section IV the main results are described, to finish in section V with the main conclusions of this study.


\section{Mathematical model}

\subsection{Flow solver} \label{sec:Flow-model}

The compressible version of the Reynolds Averaged Navier-Stokes equations (RANS) are used to model the flow, closing the mathematical system with the Wilcox $k-w$ turbulence model. These set of equations can be written in conservative form as:

\begin{equation}\label{eq:NSvol}
\frac{\partial}{\partial t} \int_\Omega \mathbf{q} d\Omega=-\int_{\partial \Omega} {\rttensor{F}} \cdot \mathbf{n} dS,
\end{equation}
where vector\footnote{Bold variables are used to refer vectors.} $\mathbf{q}$ comprises the conservative variables (density, momentum and energy) and turbulent quantities, while $\Omega$ is a control volume with boundary $\partial \Omega$ and outer normal $\mathbf{n}$. The specific heat capacities of the gas at constant volume and pressure are both assumed constant, so, consequently, it can be also defined the constant adiabatic coefficient of the ideal gas. A finite volume approach is selected for the spatial discretization, namely the change of the flow conditions in a control volume $\Omega$ is given by the normal component of the flux through the control volume boundary $\partial \Omega$.
The flux density tensor $\rttensor{F}$ can be decomposed along the three Cartesian coordinate directions and comprises the inviscid, viscous and turbulent fluxes. Details of the RANS formulation can be found elsewhere, and, for the particular case of the $k-w$ turbulence model in Wilcox\cite{Wilcox2008}.

The finite volume DLR TAU-Code\cite{taucode} was used to obtain a steady state solution of the base flow equations. The flow domain $\Omega$ is discretized using a dual mesh into a finite number of subdomains $\Omega_{i}, i=1 \ldots N$, where each subdomain contains $\mathcal{N}_f$ faces. This approach is well suited to unstructured grids, using an edge-based data structure to optimize the memory requirements and computational performance of the solver. The time-accurate three-dimensional Navier-Stokes equations are marched in time towards steady state by a backward Euler implicit scheme, solved approximately by a LU-SGS (\textit{Lower-Upper Symmetric-Gauss-Seidel Method}) iterations procedure. Local time stepping and multigrid algorithms were adopted to accelerate convergence, allowing to converge a forced-steady solution of the RANS equations from flow configurations that would otherwise develop an unsteady behavior if run with an unsteady solver.

Thus, following the method of lines, the spatial discretization of system (\ref{eq:NSvol}) gives rise to a system of ordinary differential equations, that can be written in general form for a subdomain $\Omega_{i}$ as:

\begin{equation}\label{eq:dW_R}
{|\Omega|_i}\frac{\partial \mathbf{q}_i}{\partial t} + \mathbf{R}_i=\mathbf{0} \;\;\;, \qquad \mathbf{R}_i=\sum_{j=1}^{\mathcal{N}_f} {\rttensor{F}}_j \mathbf{n}_j, \qquad i=1 \ldots N
\end{equation}
where $\mathbf{R}_i$ is the residual in subdomain $\Omega_{i}$, equivalent to the flux contributions to this subdomain, and $\mathbf{q}_i$ represents a discretized vector state solution in subdomain $\Omega_{i}$. Vectors $\mathbf{q}_{i}$ and $\mathbf{R}_{i}$ have dimensions that depend on the number of fluid variables $N_{v}$ considered. Namely, $N_{v}=$ 4, 5, 6 or 7 depending whether the problem is laminar or turbulent, two or three-dimensional. The number of subdomains depends on the number of elements of the computational mesh.

Likewise, non-slip boundary conditions and adiabatic wall are imposed on the body surface as:

\begin{equation}\label{eq:BC}
u=v=w=0  \qquad  \frac{\partial T}{\partial \mathbf{n}}=0
\end{equation}
where $\mathbf{n}$ is the normal direction to the body surface and $T$ stands for the temperature. At the external boundaries, a farfield boundary condition is used. The convective fluxes crossing the farfield boundary faces are calculated using the Advection Upstream Splitting Method (AUSM) Riemann solver\cite{liou1993} and the flow conditions outside the boundary faces are determined employing Whitfield theory\cite{Anderson1995}. Free-stream turbulence intensity on the farfield boundaries was set to 0.1$\%$, and free-stream ratio of eddy to laminar viscosity equal to $\mu_{t\infty}/\mu_\infty = 0.001$.


\subsection{Stability analysis} \label{sec:stab_theory}

Stability analysis studies the growth or decay of perturbations superimposed onto a usually steady solution of the Navier-Stokes (NS) equations. The analysis can identify which particular features are prone to evolve under modifications of the flow conditions, either by introducing a perturbation or caused by a modification of some physical or geometrical parameter. The growth of these features would give rise to a completely different flow configuration. Examples of application of flow stability to fluid dynamics problems can be found in the literature for a large variety of flow topologies\cite{Chomaz2005, Theofilis2011, Luchini2014, Taira2017, Meseguer2014, Gonzalez2011}.

Following linear stability theory, any flow quantity can be decomposed into a steady or time-periodic component, upon which small perturbations are allowed to develop. We can express, without loss of generality, that:

\begin{equation}\label{eq:FV5}
\mathbf{q}(\mathbf{x},t)=\mathbf{Q}(\mathbf{x})+\varepsilon \mathbf{\widetilde{q}}(\mathbf{x},t),
\end{equation}
where $\varepsilon\ll 1$ and $\mathbf{Q}$ is the so called \textit{base flow}, point of equilibrium of Eq. (2). The small perturbations are modeled as the term $\mathbf{\widetilde{q}}(\mathbf{x},t)$.

This decomposition is introduced into the NS equations, Eq (\ref{eq:dW_R}), and a Taylor series expansion is performed around the point of equilibrium $\mathbf{Q}$. Neglecting terms of order $\varepsilon^2$ and assembling all the unknowns in vector/matrix notation, the new system in perturbed variables is obtained:

\begin{equation}\label{eq:FV6}
    \varepsilon \mathbf{B}\pdv{\mathbf{\widetilde{q}}}{t} = 
    \mathbf{R}(\mathbf{Q}+\varepsilon \mathbf{\widetilde{q}}) \approx 
    \mathbf{R}(\mathbf{Q})+\varepsilon \qty[\pdv{\mathbf{R}}{\mathbf{q}}]_{\mathbf{Q}}\mathbf{\widetilde{q}}.
\end{equation}
where the diagonal matrix $\mathbf{B}$, with leading dimension $N_{v} \times N$, contains the volumes associated to each cell and $ \qty[\pdv*{\mathbf{R}}{\mathbf{q}}]_{\mathbf{Q}}$ represents the jacobian of the fluxes evaluated in the base flow. The stability of the system is therefore governed by the properties of the last.
If the \textit{base flow} $\mathbf{Q}$ is steady, the linear system (\ref{eq:FV6}) is autonomous and its coefficients are independent of $t$. 

If a Fourier expansion is considered in time ($t$) and spanwise ($z$) direction, the following decomposition holds:
\begin{equation}\label{eq:ansatz0}
	\mathbf{\widetilde{q}}(\mathbf{x},t) = \mathbf{\widehat{q}}(x,y)e^{i(\beta z - \omega t)} + c.c.
\end{equation}
where $\omega$ is a complex scalar, $\beta$ represents the spanwise wavenumber, $\mathbf{\widehat{q}}$ describes the complex disturbance function of $x$ and $y$, and $c.c.$ stands for complex conjugate. Complex conjugation is included since the pair $( \mathbf{\widehat{q}}(x,y),\omega)$ and its complex conjugate are both solutions of the linearized NS equations, while $\mathbf{\widetilde{q}}$ is real.\\
In pure two dimensional cases, as the one here analyzed, the velocity component and all derivatives in the spanwise direction are neglected for both \textit{base flow} and disturbances, and $\beta$ is set to zero. Renaming $- i \omega \rightarrow \omega$, the final decomposition is finally considered:  

\begin{equation}\label{eq:ansatz1}
	\mathbf{\widetilde{q}}(\mathbf{x},t) = \mathbf{\widehat{q}}(x,y)e^{\omega t} + c.c.,
\end{equation}

Substitution of the \textit{ansatz} given by equation (\ref{eq:ansatz1}) into the linear system (\ref{eq:FV6}) and considering $\mathbf{R}(\mathbf{Q})=0$ (obtained by the steady RANS computation), equation (\ref{eq:FV6}) is transformed into a two-dimensional generalized eigenvalue problem:

\begin{equation}\label{eq:GEV}
   \left[\frac{\partial \mathbf{R}}{\partial \mathbf{q}}\right]_{\mathbf{Q}}\widehat{\mathbf{q}}=\omega \mathbf{B} \widehat{\mathbf{q}},
\end{equation}
which can also be expressed as:

\begin{equation}\label{eq:GEV2}
    \mathbf{A}\widehat{\mathbf{q}}=\omega \mathbf{B} \widehat{\mathbf{q}}.
\end{equation}
with $\mathbf{A}=\left[\frac{\partial \mathbf{R}}{\partial \mathbf{q}}\right]_{\mathbf{Q}}$ being the linearized Jacobian matrix, and $\omega=\omega_r+i\omega_i$ the complex eigenvalues of the generalized system. According to this formulation, the real part $\omega_r$ is referred as the amplification rate of the corresponding eigenmode, with the imaginary part $\omega_i$ being the pulsation, related to the associated frequency as $St = \omega_i/2\pi$. If the real part of the eigenvalue takes a positive value, the associated eigenmode will eventually become dominant over the steady \textit{base flow}, whereas a negative real part will indicate a damping behavior of the associated eigenmode.

The DLR TAU-Code includes a linear frequency domain solver, which uses a analytic linearized jacobian matrix. This capability notably simplified the computational effort to obtain the input needed for the stability analysis.


\section{Numerical procedure}

\subsection{Base flow solver} \label{sec:flow-solver}

\begin{figure}
	\centering	
	\includegraphics[{width=0.5\textwidth}]{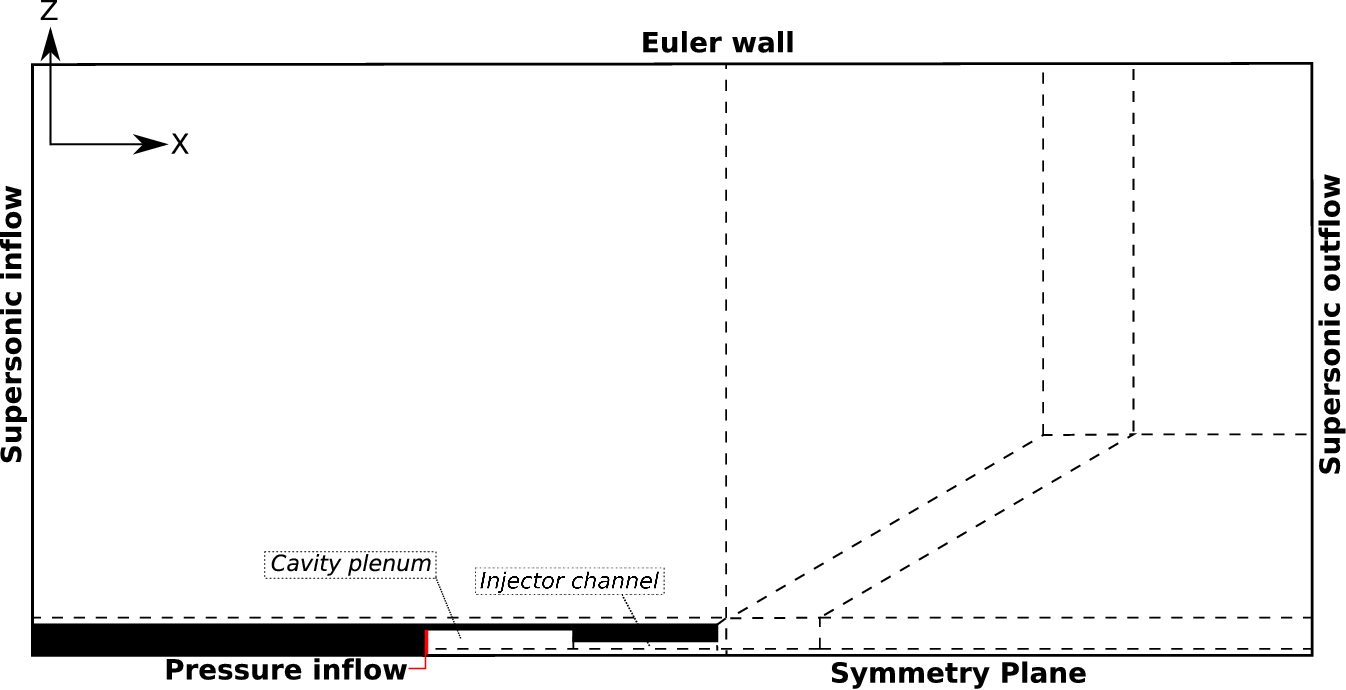}
	\caption{Geometry definition (solid) and grid topology (dashed-lines) used on the mesh generation. Only half domain is shown for simplification.}
	\label{f:geometry}
\end{figure}

The domain of reference of the problem (see Figure \ref{f:geometry}) had been previously studied by Saracoglu \cite{Saracoglu2013} and Gorbachova \cite{Gorbachova2016}. The former used extensive URANS simulation to give a comprehensive analysis of the effects of the trailing edge purge flow on the flow topology, and the latter did a first approach to the problem throughout global stability analysis. The geometrical approximation was a simplification of the industrial problem, where only the details of the flow in the vicinity of the trailing edge of a turbine blade were studied. The geometry also included the internal cooling passages, composed by a cavity plenum, where the purge flow was ejected, and the injector pipe, directly communicating the plenum with the base region. 

The computational domain was limited by straight boundaries, with a length of 21.75$\times$\textit{d} (with \textit{d} as the trailing edge thickness, equal to 0.02 m) and a width of 20$\times$\textit{d}. The injector pipe had a length and width of 2.5$\times$\textit{d} and 0.3$\times$\textit{d}, respectively. To keep subsonic conditions inside the cavity plenum, its width was kept three times larger than the injector pipe. The characteristic geometrical lengths of the analysis are shown in Table \ref{geometry_values}.

Boundary conditions of the control volume were defined as follows: Supersonic inflow and output, Euler wall and symmetry plane were set at left, right, upper and lower limits respectively; non-slip wall was used for all internal surfaces and a reservoir-pressure inflow boundary condition was set to simulate the input of the cooling in the cavity plenum area (Fig. \ref{f:geometry}).
 
The Mach number at the inflow boundary was set to 1.5, keeping a total temperature of 365 K and a Reynolds number of $9.4 \times 10^6$, based on the trailing edge length (11.65$\times$\textit{d}). The flow rate at the cooling flow exit, from now on referred as the \textit{blowing rate}, was defined by the cooling flow static temperature, kept to 300 K, and its total pressure, expressed as a percentage of the free stream flow total pressure.

\begin{figure}
	\begin{subfigmatrix}{2}
        \subfigure[Computational mesh]
	       	{{\includegraphics[width=0.45\textwidth]{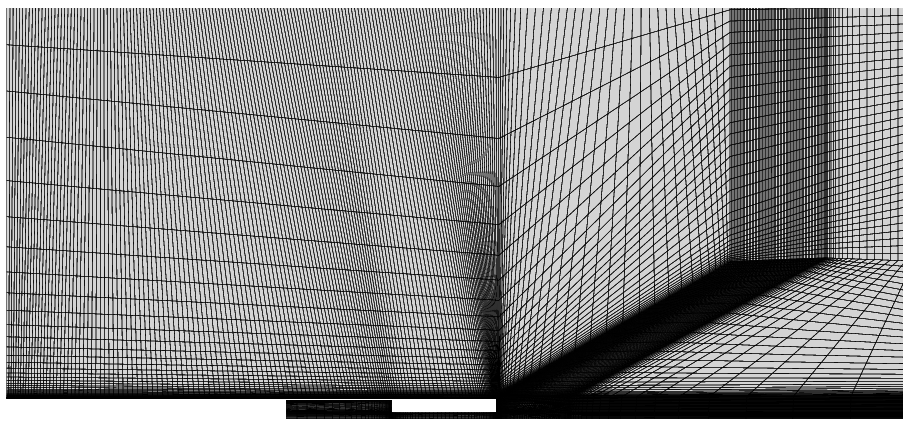}}}
		 \subfigure[Details of the mesh in the of the cavity and channel area]
	       	{{\includegraphics[width=0.45\textwidth]{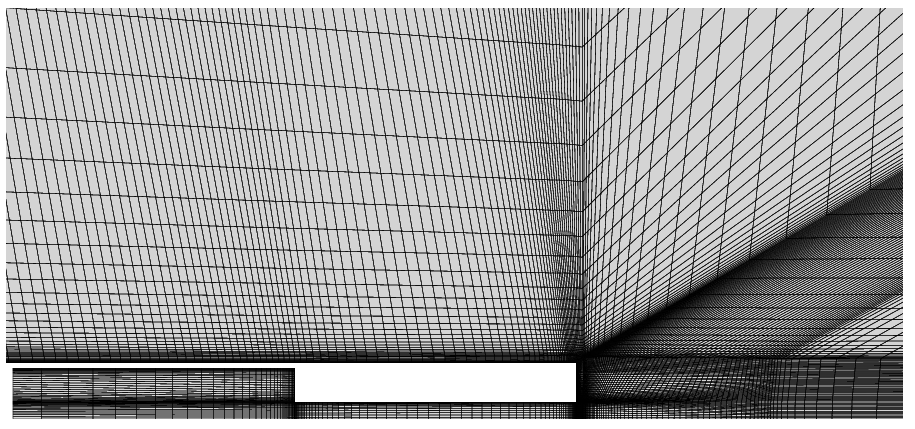}}}
	\end{subfigmatrix}
	\caption{Computational mesh. Only a quarter of total nodes are represented for clarity.}
	\label{f:small_multiple}
\end{figure}

\begin{table}[hbt!]
\caption{Fluid domain dimensions}\label{geometry_values}
\centering
\begin{tabular}{lc}
\hhline{==}
Boundary & Length  (m) \\
\hline \hline
Domain length & 0.4346 \\
Domain height & 0.4 \\
\hline
Trailing edge length & 0.233 \\
Trailing edge  height & 0.02 \\
\hline
Cavity plenum length & 0.05 \\
Cavity plenum height & 0.018 \\
\hline
Injector channel length & 0.05 \\
Injector channel height & 0.006 \\
\hhline{==}
\end{tabular}
\end{table}
Since the study aimed to understand the different flow features present in the flow and, in particular, the nature of the non-symmetry behavior, a global stability analysis over the steady symmetric base flow was performed. Thus, only half of the domain was simulated using a RANS approach, imposing a symmetry boundary condition on the plane of symmetry to obtain a steady solution. After obtaining a ``half" base flow, mesh and solution were mirrored into the symmetry plane and the stability analysis of the two-dimensional full domain was carried out. The global analysis of the mirrored-full domain around this symmetric flow at different blowing ratios will show non-symmetric modes in the analysis, which could eventually grow and generate a new non-symmetric flow configuration. To clarify the exposition, when the term \textit{half-domain} is used on this text, will refer to the mesh of Figure \ref{f:small_multiple}, with a symmetry plane on its lower boundary; on the contrary, when \textit{full-domain} is called, it will refer to simulations done on a mirrored mesh, considering upper and lower surfaces of the trailing edge. Similar methodology on the detection of flow bifurcation can be found on the literature on the detection of global modes of symmetric sudden expansions\cite{Fani2012} or X-junction flow configurations\cite{Lashgari2014}.

The domain was discretized using a quad-structured mesh, generated using an O-Grid topology and dividing the geometry into regions of interest. Wall normal grid distances were set small enough so values of $y^{+}$ lower than 1 were kept on the body surfaces, and all the scales of the boundary layer were fully resolved without the use of any wall model. Grid topology and mesh details are pictured on Figures \ref{f:geometry} and \ref{f:small_multiple}. The mesh spatial resolution was concentrated at the contraction-expansion and on the shock wave areas.

To ensure mesh independent results, a mesh convergence study was performed for a blowing rate of 18\% for a half-domain configuration. Six different meshes of increasing number of elements were defined, named from M1 to M6. A systematic refinement was performed to evaluate the flow sensitivity to the mesh density. At each step the mesh was refined increasing $n$ times the number of nodes, taking the mesh M1 as the reference. This refinement law preserved the topology of the mesh and guaranteed that Richardson extrapolation formulas were applicable to obtain an accurate rate of convergence\cite{Fraysse2013}. The convergence criteria, evaluated through the global density residual $res^n_\rho$, was set up to values lower than $10^{-6}$. It was defined as:

\begin{equation}
    \|{res^n_\rho}\Arrowvert = \sqrt{\sum^{N}_{j=1} \frac{[res^n_\rho(j)]^2}{N}}
\end{equation}
where $N$ denoted the number of grid points.

Detailed results from the mesh convergence study are shown in Table \ref{meshconv}. At each mesh the pressure and density values at the trailing edge were monitored. According to these results, the mesh M4 was selected as a good compromise between accuracy and computational efficiency.

\begin{table}
	\caption{Mesh Convergence Analysis performed for a blowing rate of 18\% for a half-domain configuration. $P_b/P_b^*$ represented the relation between the pressure with and without purge flow applied, and $\rho_{purge}/\rho_f$ was the relation between the density values at the cavity plenum and at the base region.}\label{meshconv}
	\centering
	\begin{tabular}{lccc}
		\hhline{====}
		Mesh & Number of nodes & $P_b/P_b^*$ & $\rho_{purge}/\rho_f$ \\
		\hline
		M1   & 12010          & 1.26932202     &  1.03157676 \\
		M2   & 48260        & 1.29883128     &   1.03038449 \\
		M3   & 104630       & 1.30545566    &   1.02675552 \\
		M4   & 207106       & 1.31826117      &   1.01727924 \\
		M5   & 321090       & 1.31929642    &   1.01407829 \\
		M6   & 424640      & 1.31858961     &   1.01426252 \\
		\hhline{====}
	\end{tabular}
\end{table}


\subsection{Stability analysis} \label{sec:stab-method}

The real non-symmetric operator $\mathbf{A}$ (Eq. \ref{eq:GEV}) was extracted using the first-discretize-then-linearize technique implemented in the DLR-TAU solver, and saved using a compressed sparse row format. For the particular case of mesh M4, the leading dimension of the matrix $\mathbf{A}$ was approximately $1.2 \times 10^{6}$, and the number of non-zero elements was $9.6 \times 10^{7}$. The real diagonal operator $\mathbf{B}$, representing the volume of each element, was also obtained from the solver.

For each eigenvalue there is an associate eigenvector that represents the underlying physical feature (density, momentum in the spatial directions ($x,y,z$), energy and turbulent viscosity), which will eventually decay or grow depending on the sign of the real part of the eigenvalue. To calculate the eigenvalues and eigenvectors of system \ref{eq:GEV2}, the Implicitly Restarted Arnoldi Method (IRAM) algorithm implemented in the ARPACK library\cite{ARPACK} was used. The Arnoldi algorithm is a good technique for approximating the eigenvalues of large sparse matrices, but however only the eigenvalues with larger module can be obtained by a straightforward application of the algorithm. To capture the region of the complex plane where the unstable eigenvalues layed, a shift-and-invert preconditioning technique\cite{Saad2011} was used. Not all the eigenvalues were computed, but only a finite number close to an area of interest, defined by the shift parameter. At each iteration of the IRAM algorithm, the shift-and-invert method requires the inversion of a linear system, which made this procedure computationally expensive. Moreover, due to the stiffness associated to the underlying compressible and turbulent equation system, the jacobian matrix $\mathbf{A}$ is very badly conditioned, so iterative methods to solve the linear system fail to reach a proper convergence. In this work, a full LU factorization was performed for the Jacobian matrix $\mathbf{A}$. This strategy, used here in a compressible finite volume context, has also been followed in stability analysis of incompressible flows in finite elements discretizations and in the context of spectral methods \cite{DeVicente2011,Ferrer2014,Browne2014} and highly compressible flows with turbulence modelling\cite{Iorio2014}. Unfortunately, the full LU decomposition scales as the cube of the number of unknowns ($(N_v \times N)^3$), becoming the bottleneck of the overall process and consuming a large amount of computational resources\cite{Iorio2014}. To partially alleviate this requirement, sparse matrix format and parallel algorithms were used. This was done through an OpenMP and MPI implementation of the algorithms and the use of the MUMPS library\cite{MUMPS}.

Additionally, in order to reduce the dimension of the eigenvalue problem, a domain reduction technique was applied on this work prior to the LU decomposition.

\subsubsection*{Domain reduction methodology}

For completeness of exposition, we shortly reproduce here the mathematical foundation of this method, additional information can be found in Sanvido et al.\cite{Sanvido2017}. 

Arnoldi method belongs to the broad category of projection methods which aim to obtain the part of the spectrum of matrix $\mathbf{A}$ where the dominant eigenpairs lay.

The domain reduction (DR) strategy can be considered, somehow, a projection technique where the original matrix is reduced to a computationally affordable one. The projection in the domain reduction procedure is a purely geometrical technique; starting from a domain of reference $\Omega_n$, defined from the computational mesh, and a vector of unknowns $\mathbf{q_n}$, comprising the variables involved in the Linear Navier Stokes equations, we consider a subdomain of the computational mesh $\Omega_m$, a region of interest, which contains $m<n$ degrees of freedom.  For simplicity, we order the unknowns related to the subdomain in the first positions, in the global vector of unknowns $\mathbf{q_n}$. Namely,
$$
\mathbf{q_n}  =\{q_1 \cdots q_n \}^T = \{ q_{\{1\cdots m\}} | q_{\{(m+1)\cdots n\}}\}^T,
$$
under this assumption, the projection of the unknowns from $\Omega_n$ to $\Omega_m$ is easily performed through the projection operator, $\mathbf{P}$ defined as:

\begin{equation}\label{proyection}
\mathbf{q_m} = \mathbf{P} \mathbf{q_n} \mbox{ with }   \mathbf{P} = \left(
\begin{array}{ccccc}
1 & 0 & 0 & \cdots & 0 \\
0 & 1 & 0 & \cdots & 0 \\
0 & \vdots & \vdots & \ddots & 1 \\
0 & 0 & \cdots & 0 & 0 \\
0 & 0 & \cdots & 0 & 0 \\
\end{array}
\right)_{(n,m)},
\end{equation}
where the first $m$ rows of $\mathbf{P}$ form an identity matrix of order $m$ and the remaining $(n-m)$ rows are identically zero. It is easily checked that $\mathbf{P}$  is a orthonormal projector and its columns form a orthonormal basis on the subdomain defined by $\Omega_m$, so the projection of the Jacobian $\mathbf{A_n}$ onto this subspace is easily computed as:

\begin{equation}
    \mathbf{A_m} = \mathbf{P^T} \mathbf{A_n} \mathbf{P},
\end{equation}
where the superscript $T$ stands for the transpose and $\mathbf{A_m}$ the orthonormal projection of $\mathbf{A_n}$ in the subspace of dimension $m$.

Hence, the original Jacobian (in $\Omega_n$) can be replaced by its projection (in $\Omega_m$) and the "reduced" eigenpairs computed. Sanvido et al.\cite{Sanvido2017} showed that DR technique permits to recover the most relevant disturbances related to the specific region of interest, with the added value of filtering part of the spectra that is not relevant for that particular region. The authors proved that the approximation will be valid as long as the reduced domain contains the structural sensitivity region of the dominant eigenmodes.

\begin{figure}[!ht]
	\centering
    \begin{subfigmatrix}{2}
        \subfigure[Influence of the domain reduction technique on the eigenvalue analysis.]{\includegraphics[height=5cm,keepaspectratio]{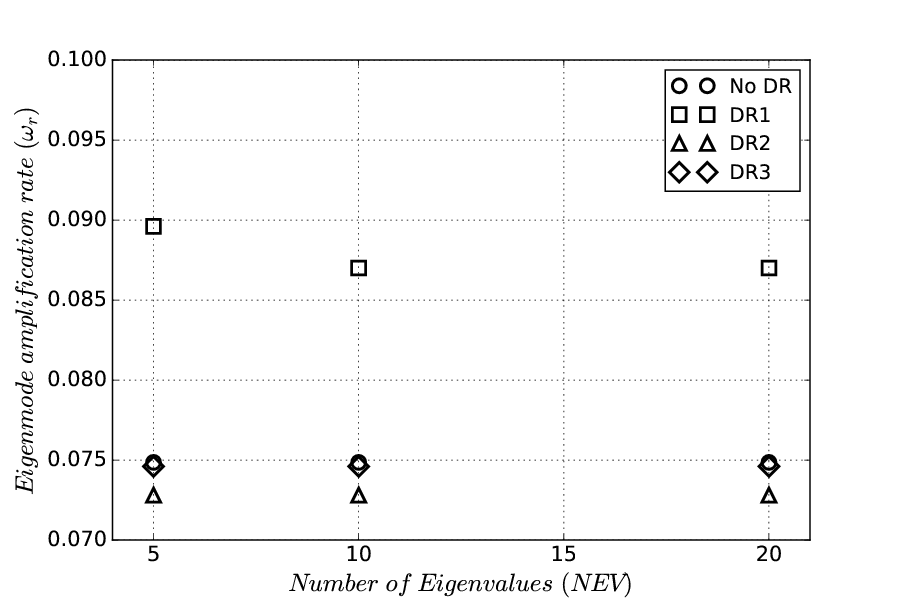}}
        \subfigure[Illustration of the different reduced domains used for the analysis.]{\includegraphics[height=5cm,keepaspectratio]{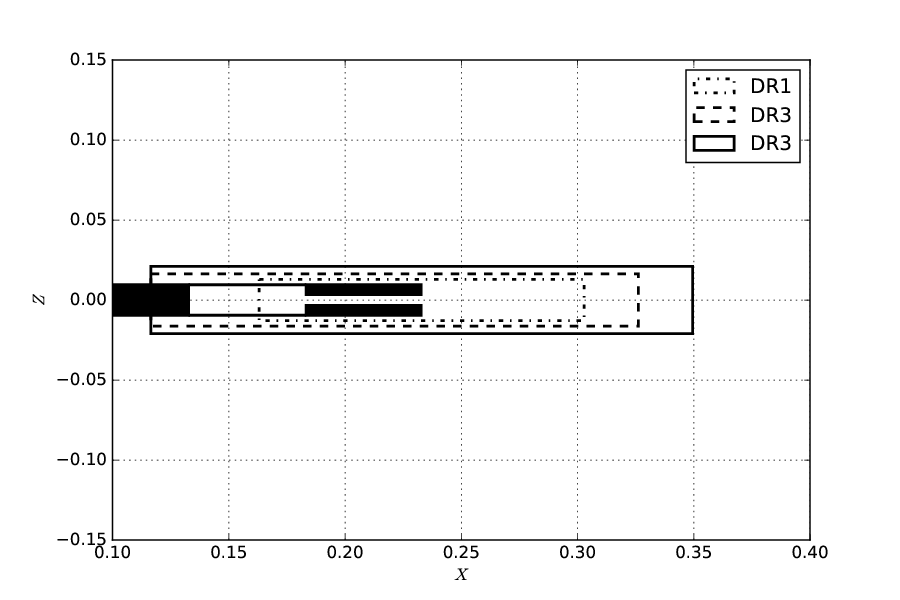}}
    \end{subfigmatrix}
    \caption{Domain reduction technique convergence analysis.}
 \label{f:num-egvconv}
\end{figure}

On the case of study, the region of interest was concentrated around the base region and the cavity-injector area (Fig. \ref{f:num-egvconv}-(b)), representing a small section of the whole computational domain. 

In order to illustrate the influence of the domain used on the stability analysis, part of the eigenspectrum for a blowing rate of 18\%  is shown in Figure \ref{f:num-egvconv}-(a). The abscissa represents the number of requested eigenvalues (NEV), set up as a third of the projected Krylov subspace size (NCV), and the ordinate shows the real part of the least stable eigenvalue, $\omega_r$. Three different domains, denoted as DR1 to DR3, were considered and compared to the results obtained without any domain reduction technique (Fig. \ref{f:num-egvconv}-(b)). The tendency of the results are in good agreement with the one predicted by Sanvido et al.\cite{Sanvido2017}, where the eigenvalue showed a strong sensitivity to the domain extension. It is not the aim of this work to explain the reason of these differences but only the applicability of the methodology. A complete description of the sources and estimation of the errors is given in Sanvido et al.\cite{Sanvido2017}. According to these results, DR3 was used for the stability computations on the rest of this work.

The number of degrees of freedom of the reduced domain DR3 was 436320, which meant a total of $\simeq 10^{11} $ elements of the Jacobian matrix, and a number of non-zero elements of approximately $3.3 \times 10^{7} $, almost a third of the non-zero elements of the original matrix. To highlight the advantages of the use of a reduced domain, statistics calculated on a local workstation equipped with 8 $Intel Core^{TM} i7-6700K$ CPUs and 32 GB of available RAM memory were collected. Prior to the application of the DR technique, the solve of the eigenvalue problem would take a total of 700 seconds to finalize, consuming more than 25 GB of RAM memory and with a number of non-zero elements of $8.7 \times 10^8$ for the factorized matrix. After the application of the domain reduction, the time to solve the eigenvalue problem was reduced to 210 seconds, demanding only 8 GB of RAM memory and with a number of non-zero elements of $2.8 \times 10^8$ for the factorized matrix.

\section{Results}

\subsection{Flow topology analysis} \label{sec:flow-topo}

\begin{figure}
	\begin{subfigmatrix}{2}
		\subfigure[Relation between the blowing rate and the density ratio.]{ \includegraphics[height=5cm,keepaspectratio]{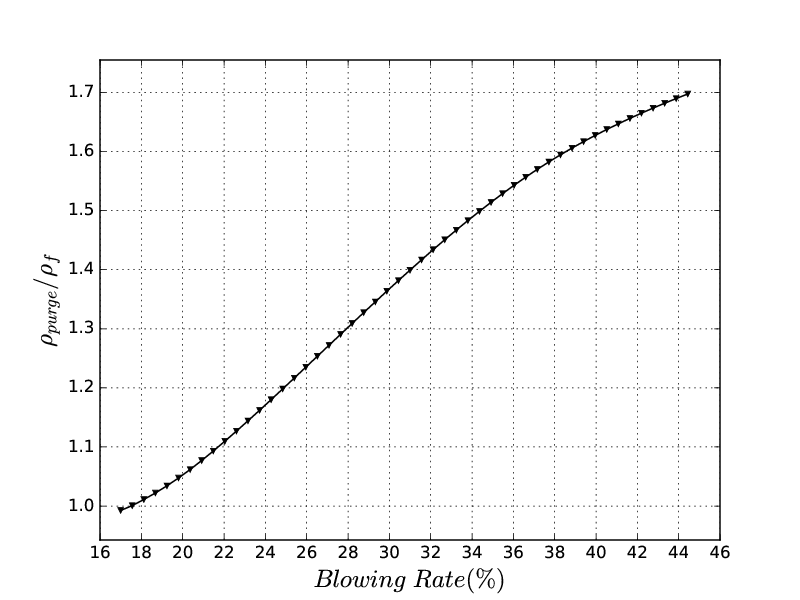}}
		\subfigure[Base pressure bifurcation, as a function of the blowing rate.]{\includegraphics[height=5cm,keepaspectratio]{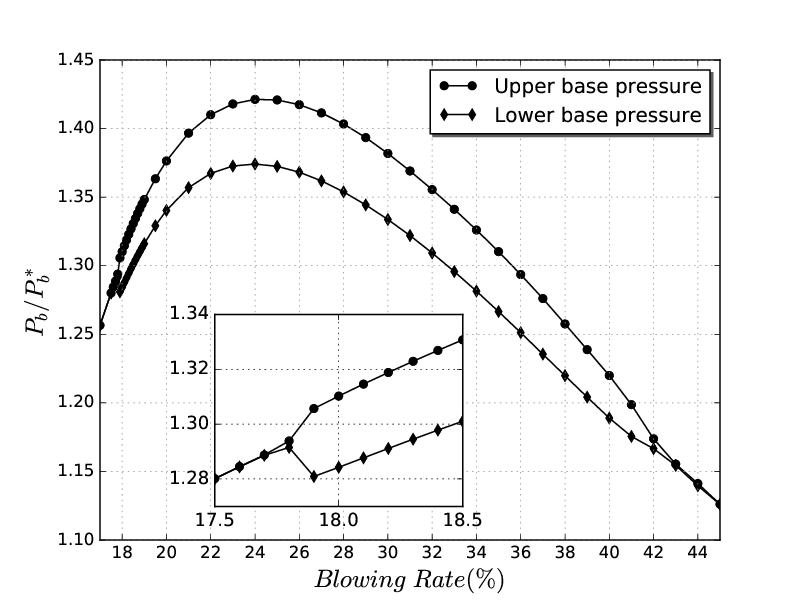}}
	\end{subfigmatrix}
	\caption{Density and Pressure values measured at the base region.}
	\label{f:baseregiondata}
\end{figure}

The results shown here are expressed as a function of the blowing rate intensity, defined as mentioned by its static temperature and total pressure, the last defined as a percentage of the free stream total pressure. Due to its widespread use in the literature related to turbomachinery trailing edge cooling, free stream to purge density ratio ($\rho_{purge} / \rho_f$) was also calculated to facilitate possible comparisons. The free stream density value is defined as the mean density on the trailing edge faces, and the density of the purge flow as the one in the cavity plenum. The relation between these two, obtained in the direct computations of the full domain, is shown in Figure \ref{f:baseregiondata}-(a). The results agreed with those obtained by Saracoglu et al.\cite{Saracoglu2013} and Gorbachova et al.\cite{Gorbachova2016}. Additionally, in Figure \ref{f:baseregiondata}-(b) the ratio between the computed base pressure at both sides of the trailing edge ($P_b$) and the base pressure for a non-blowing configuration ($P_b^*$) is shown as a function of the blowing rate. Pressure values bifurcated for a blowing rate of 17.8\%, going back to the symmetric flow state for a blowing rate of 42\% and reaching a maximum value of a 3.6\% of pressure difference at 28\% of blowing intensity. The value of the blowing rate for the maximum pressure difference did not match the maximum value for the base pressure, which took place at a blowing rate of 24\%. As it will be shown, the reason of this behavior is the appearance of a non-symmetric global mode which eventually becomes unstable and dominates the flow configuration. With the aim of capturing the instability related with the onset of this bifurcation and identifying the sources of the non-symmetry, cases where a blowing rate went from 10\% to 45\% were analyzed.

\begin{figure}
\patchcmd{\subfigmatrix}{\hfill}{\hspace{0.8cm}}{}{}
 \begin{subfigmatrix}{2}
  \subfigure[No bleeding]{\includegraphics[width=0.4\textwidth]{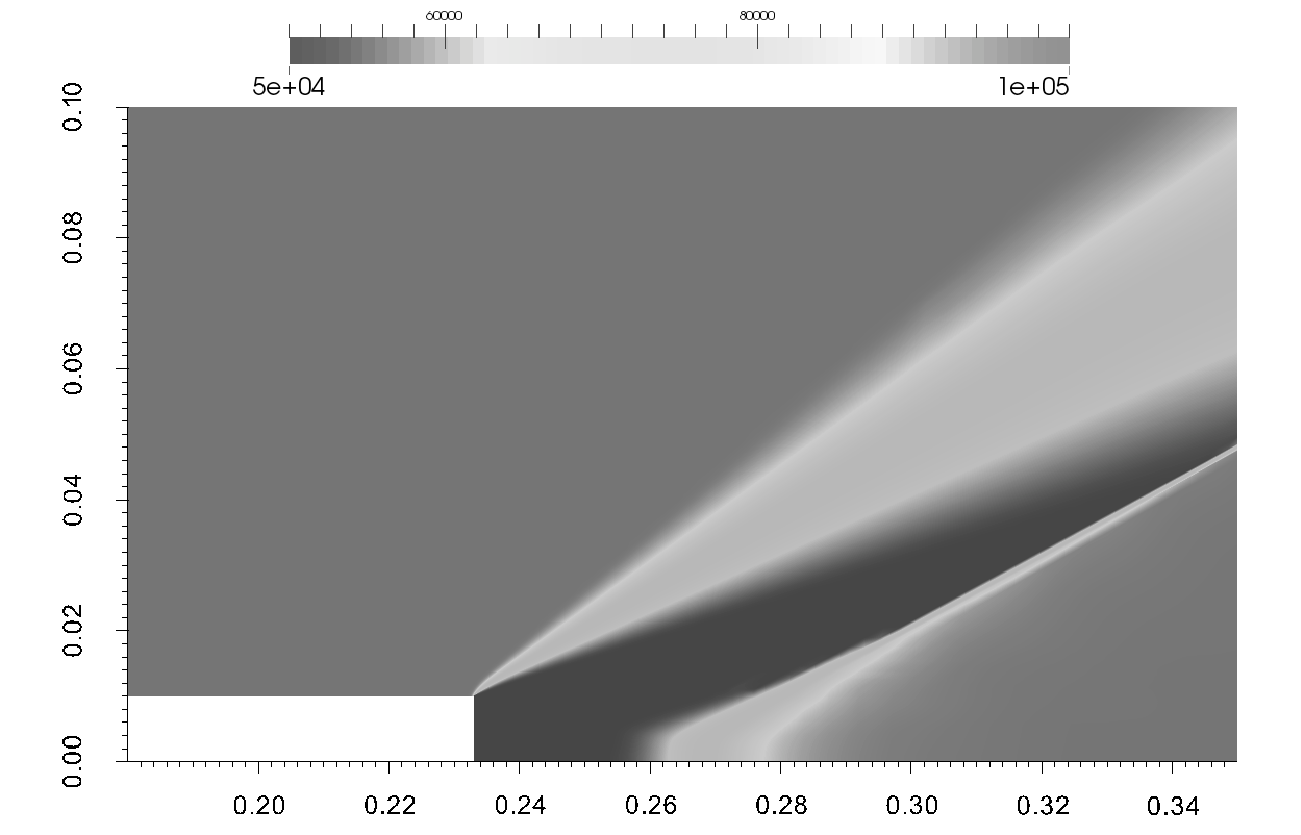}}
  \subfigure[17 $\%$ blowing rate]{\includegraphics[width=0.4\textwidth]{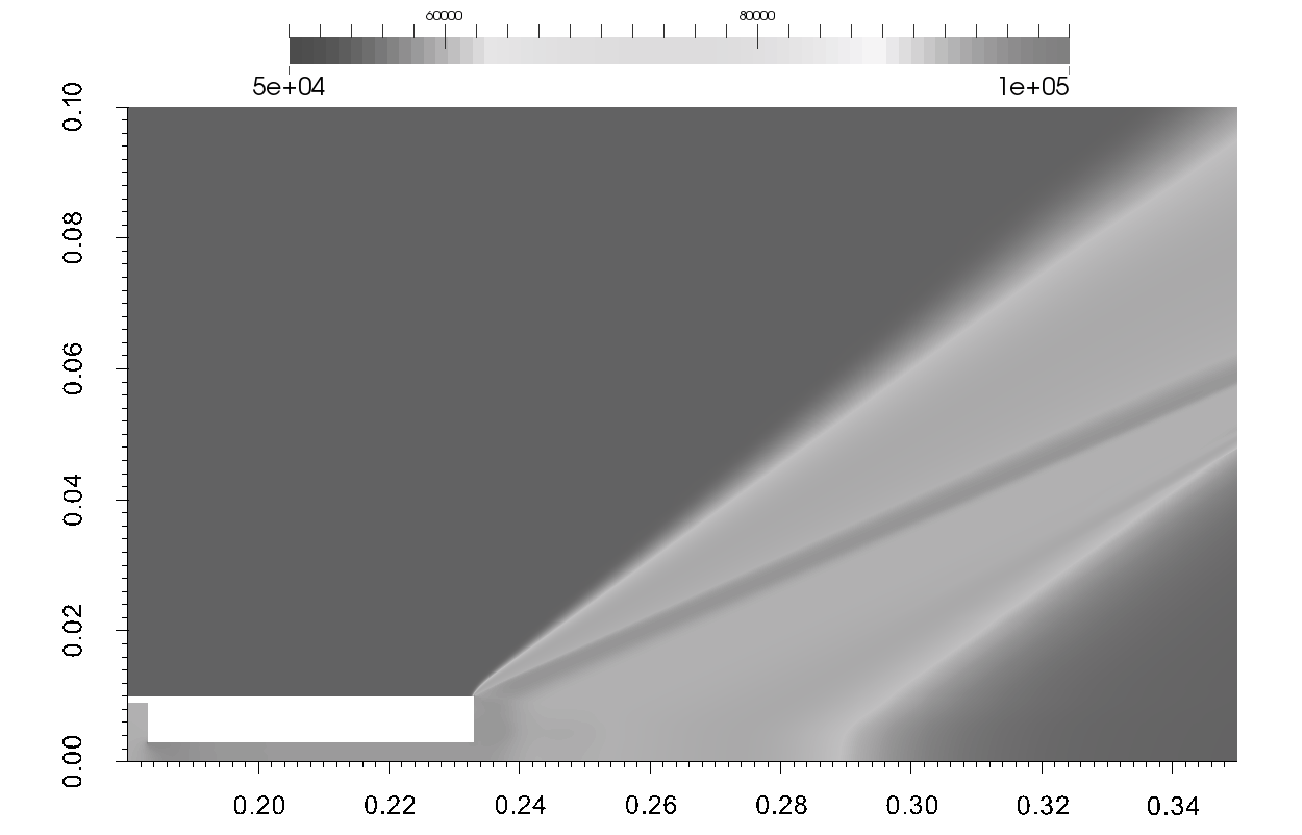}}
  \subfigure[31 $\%$ blowing rate]{\includegraphics[width=0.4\textwidth]{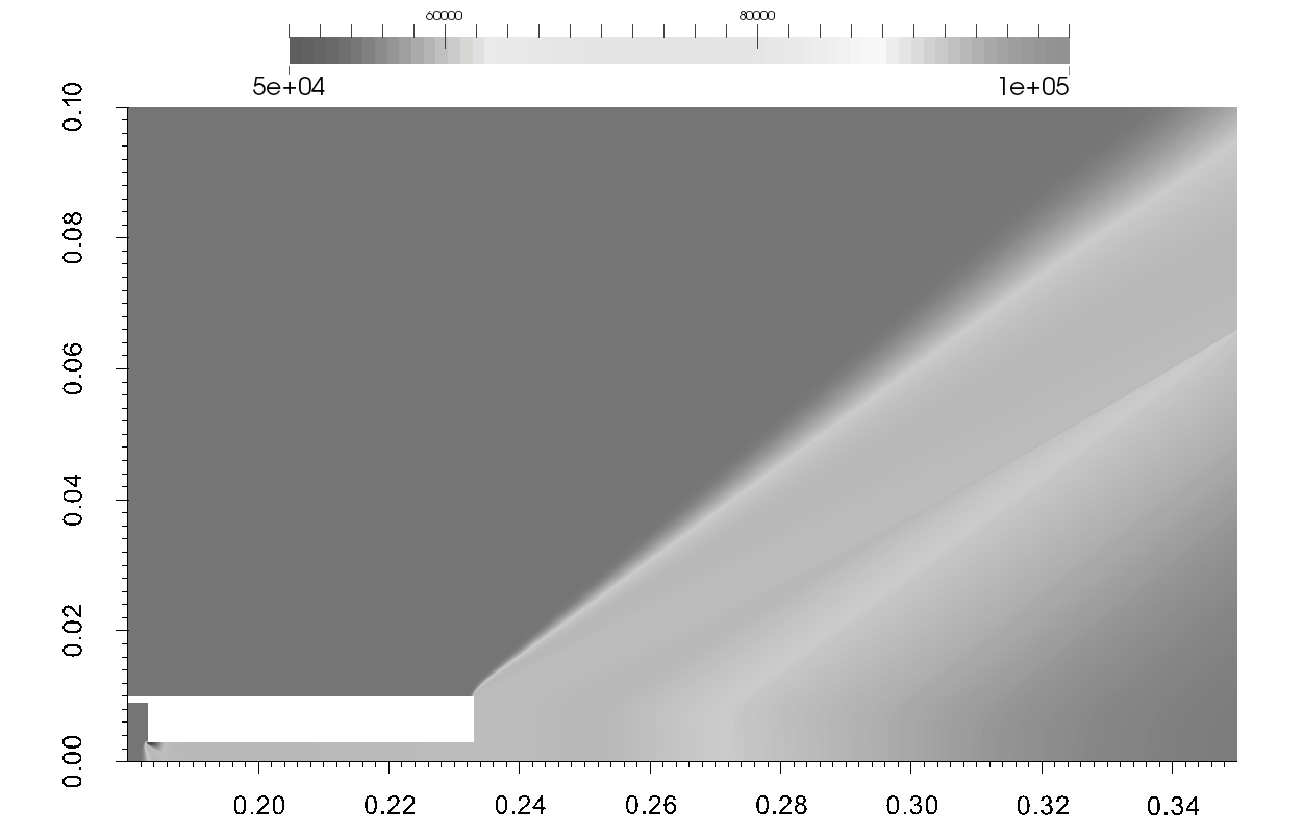}}
  \subfigure[43 $\%$ blowing rate]{\includegraphics[width=0.4\textwidth]{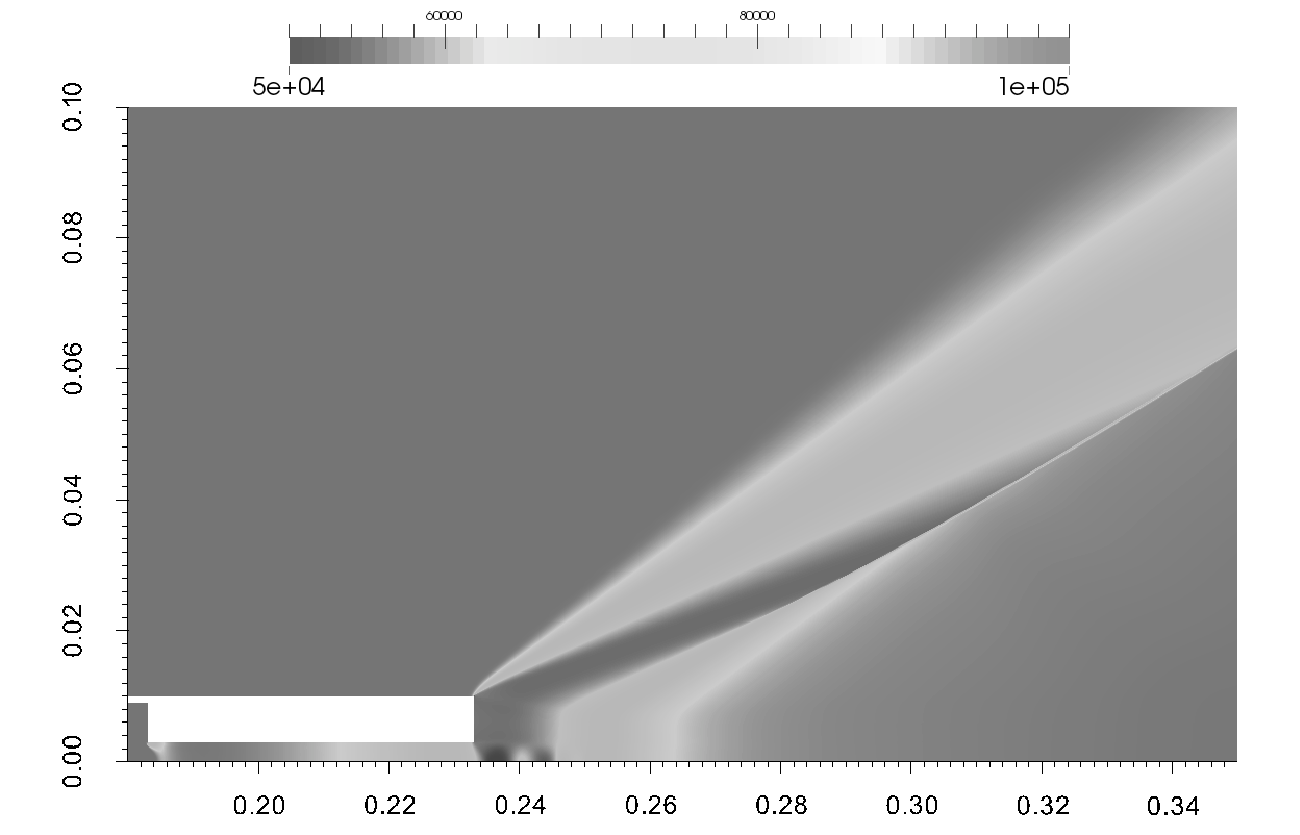}}
 \end{subfigmatrix}
 \caption{Base flow computations for different test cases and no bleeding configuration. Pressure contours.}
 \label{f:baseflow-pressure}
\end{figure}

\begin{figure}
\patchcmd{\subfigmatrix}{\hfill}{\hspace{0.8cm}}{}{}
 \begin{subfigmatrix}{2}
  \subfigure[No blowing]{\includegraphics[width=0.4\textwidth]{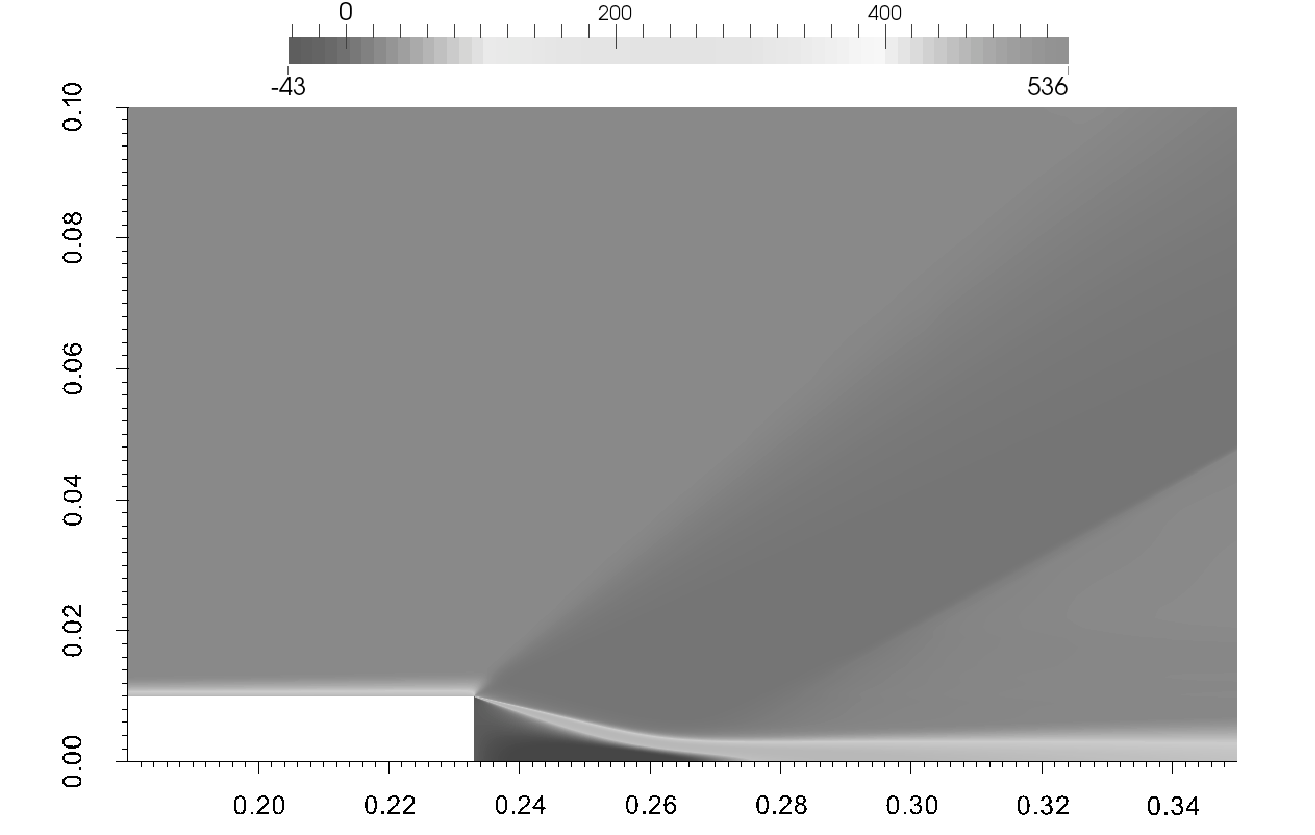}}
  \subfigure[17 $\%$ blowing rate]{\includegraphics[width=0.4\textwidth]{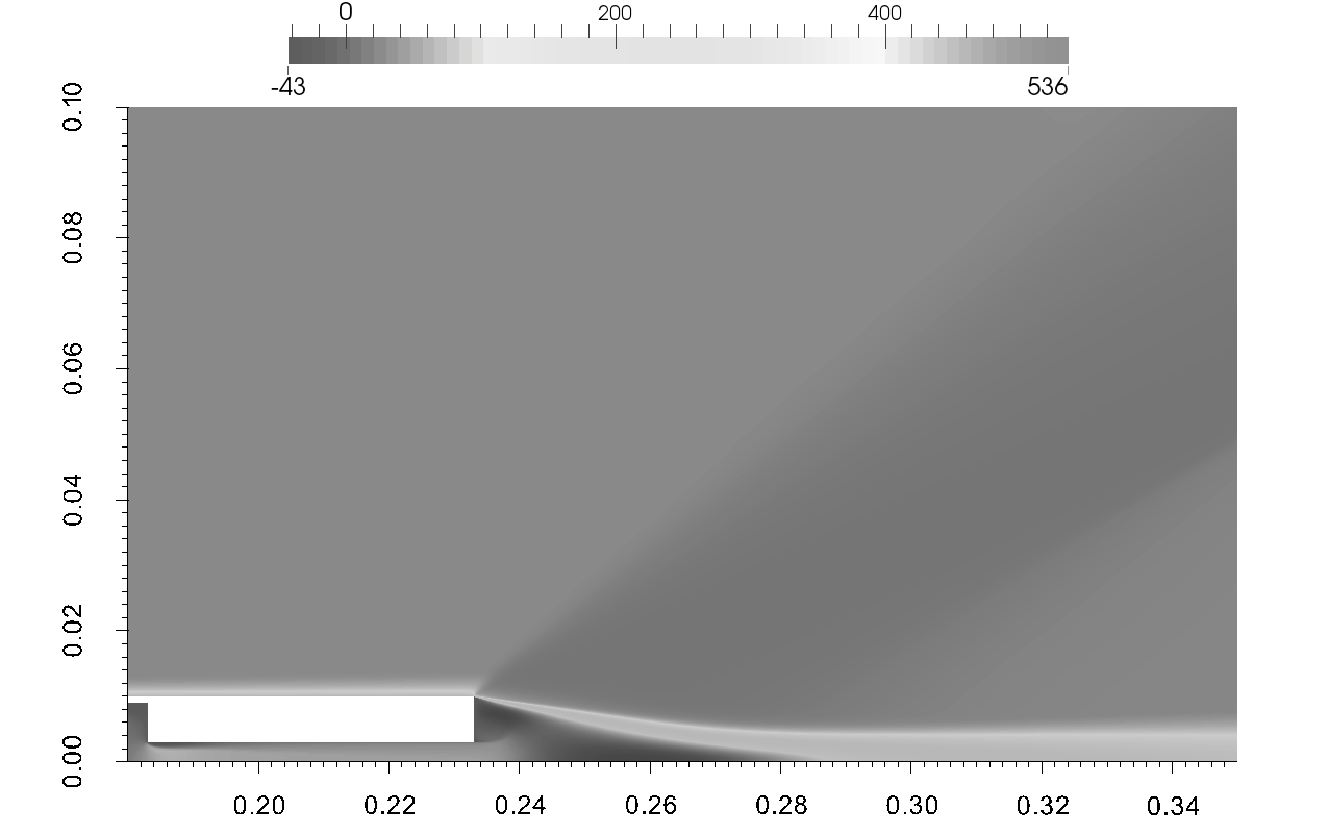}}
  \subfigure[31 $\%$ blowing rate]{\includegraphics[width=0.4\textwidth]{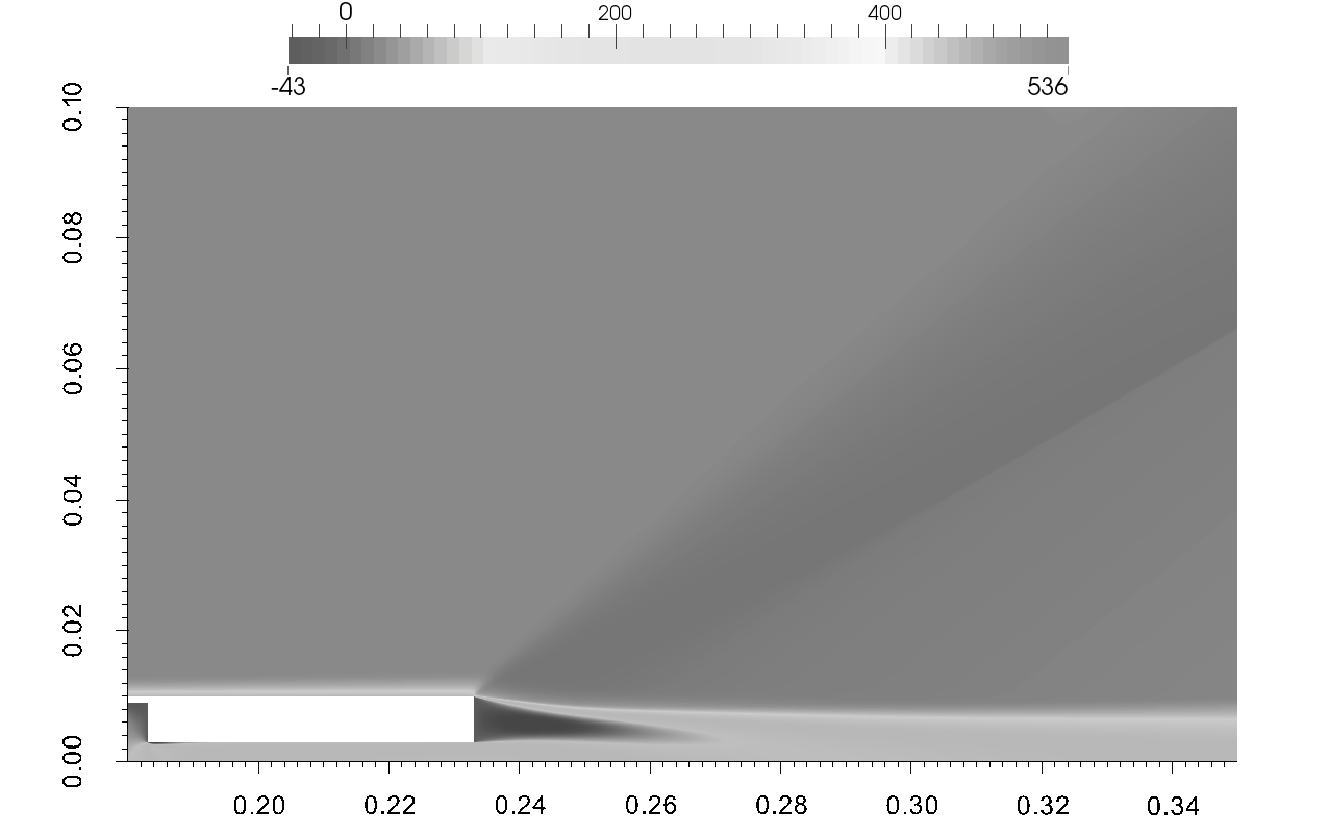}}
  \subfigure[43 $\%$ blowing rate]{\includegraphics[width=0.4\textwidth]{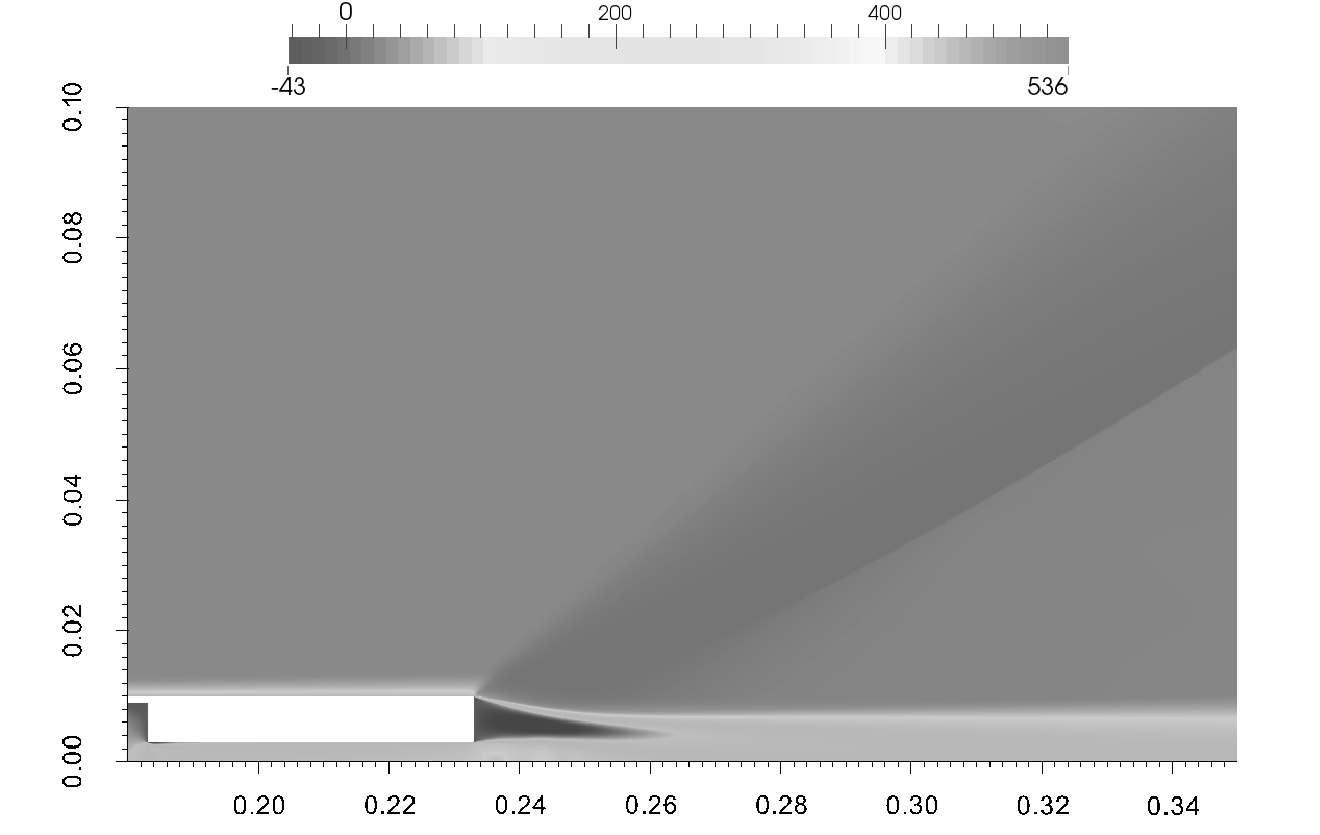}}
 \end{subfigmatrix}
 \caption{Base flow computations for different test cases and no bleeding configuration. Streamwise velocity contours.}
 \label{f:baseflow-vel}
\end{figure}

For illustration, the base flow pressure and velocity fields for some of the test cases obtained with the mesh M4, including the solution for a non-blowing configuration,  are displayed in Figures \ref{f:baseflow-pressure} and \ref{f:baseflow-vel}. These figures reveal how the shock waves decrease its intensity for low purge intensities, which combined with the filling effect of the purge jet increases the pressure of the base region. For blowing rates values higher than 24\%, the flow on the dead area starts to evacuate and the pressure at the base region decays, recovering the shock part of its original strength.

\begin{figure}
 \begin{subfigmatrix}{4}
  \subfigure[No blowing]{\includegraphics{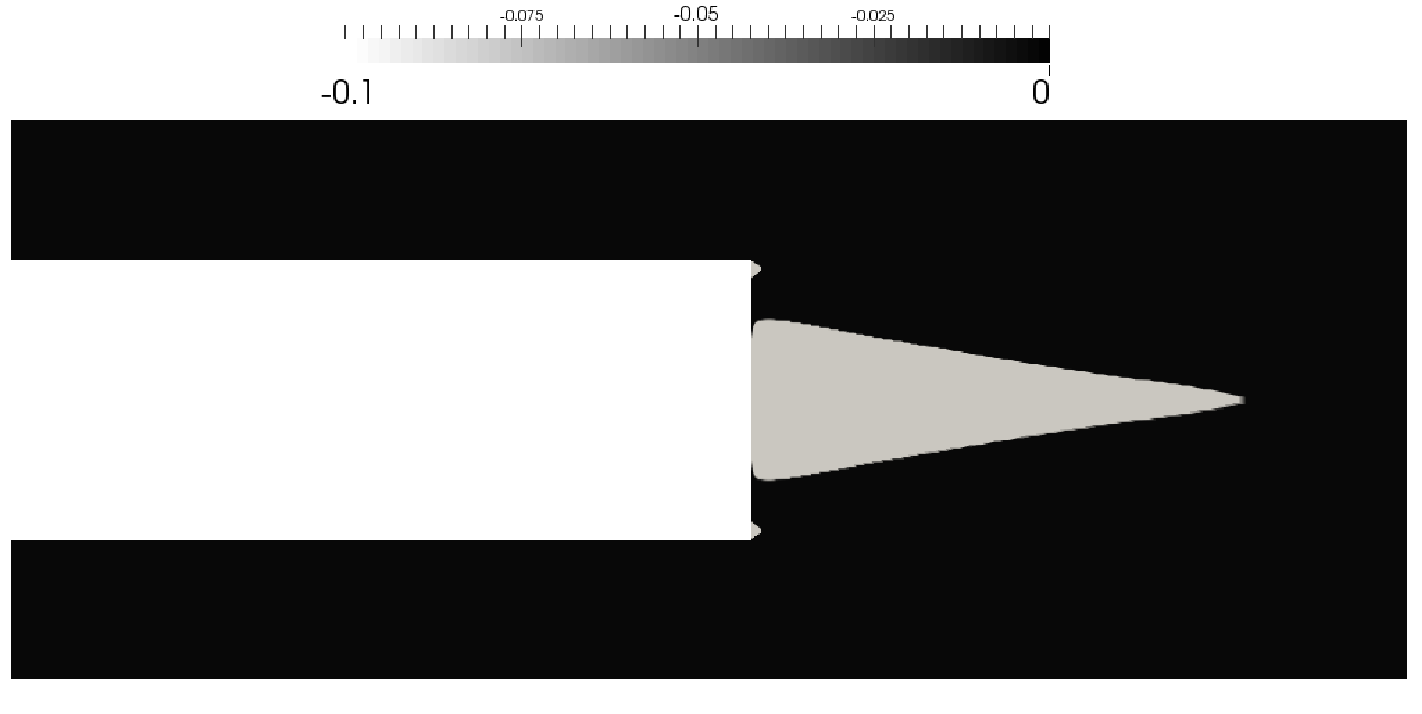}}
  \subfigure[10 $\%$ blowing rate]{\includegraphics{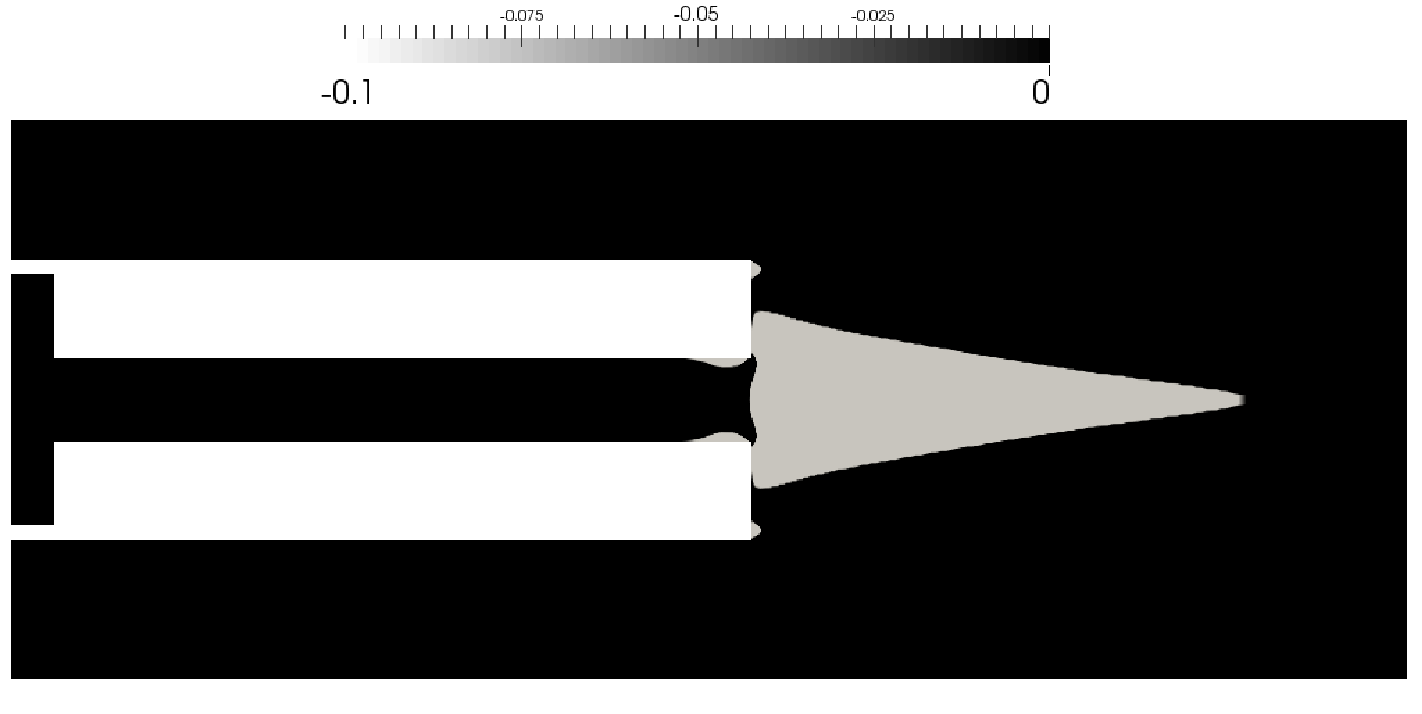}}
  \subfigure[15 $\%$ blowing rate]{\includegraphics{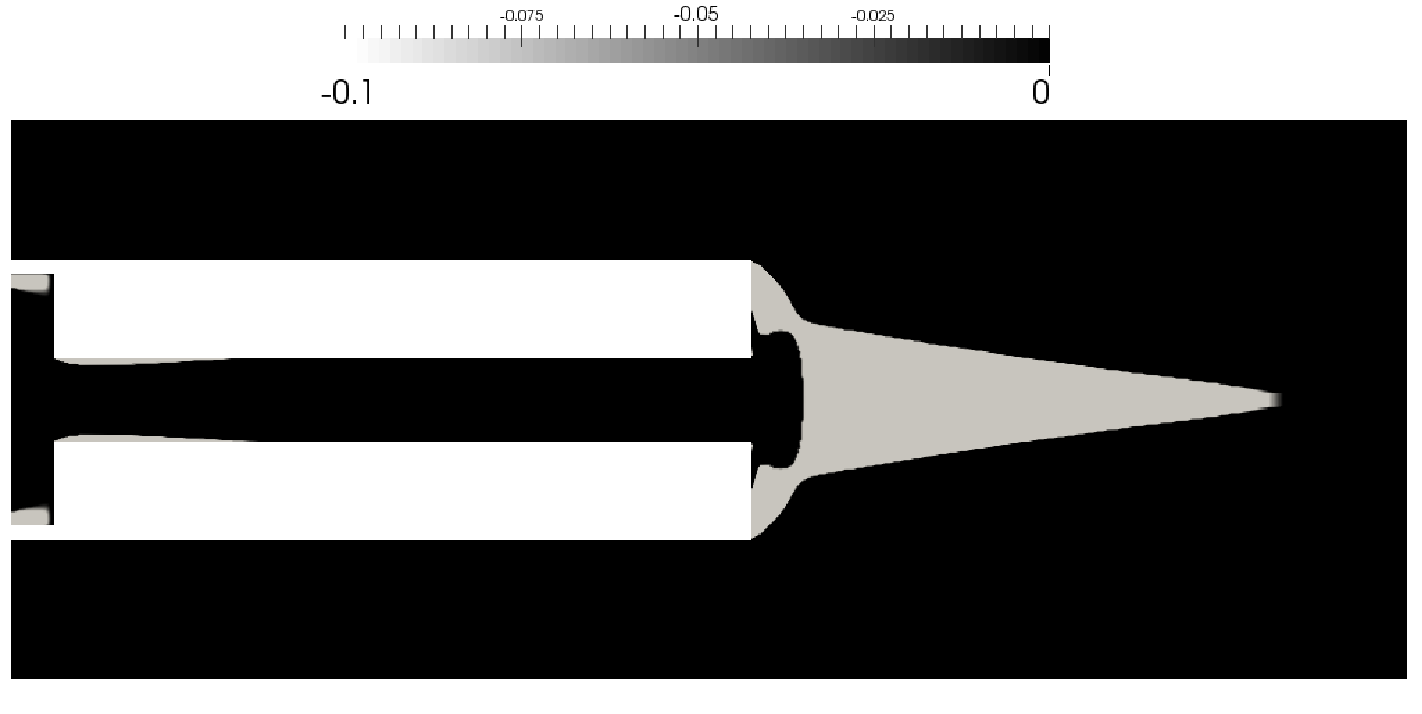}}
  \subfigure[17 $\%$ blowing rate]{\includegraphics{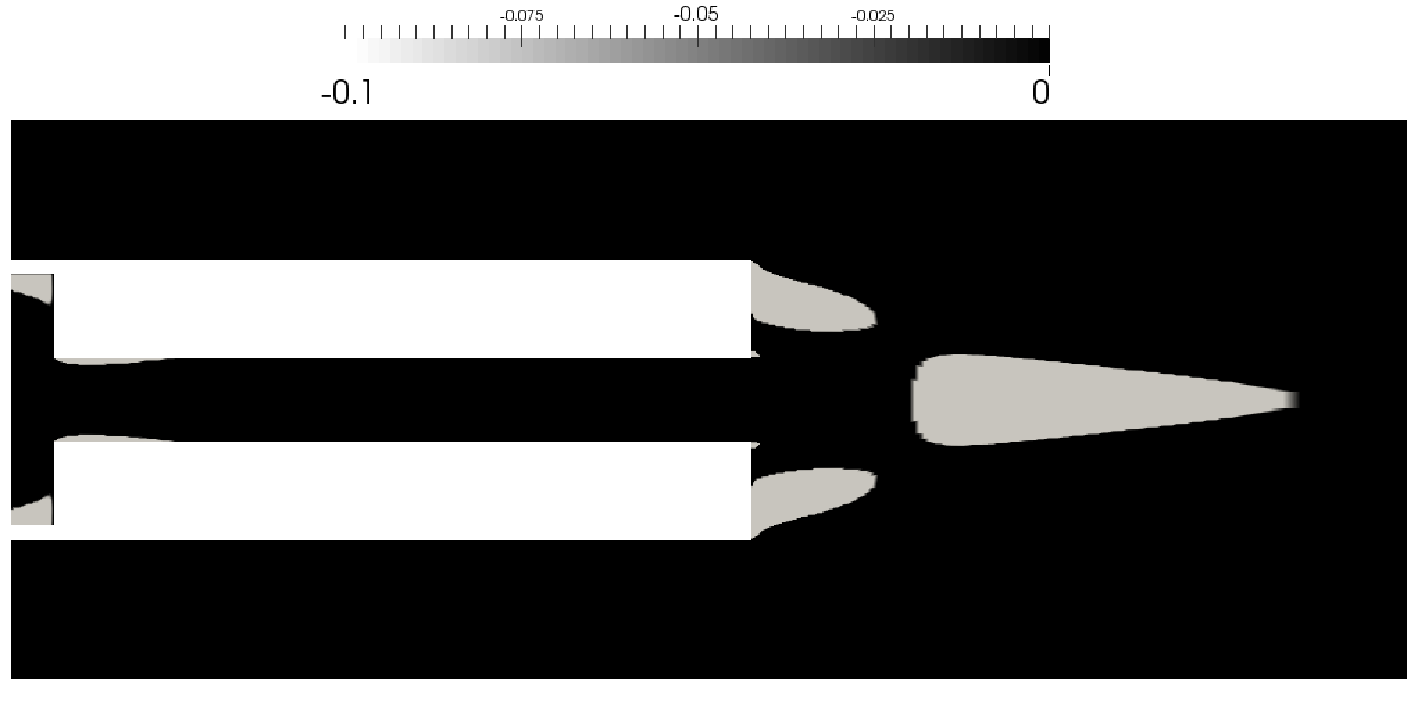}}
  \subfigure[18 $\%$ blowing rate]{\includegraphics{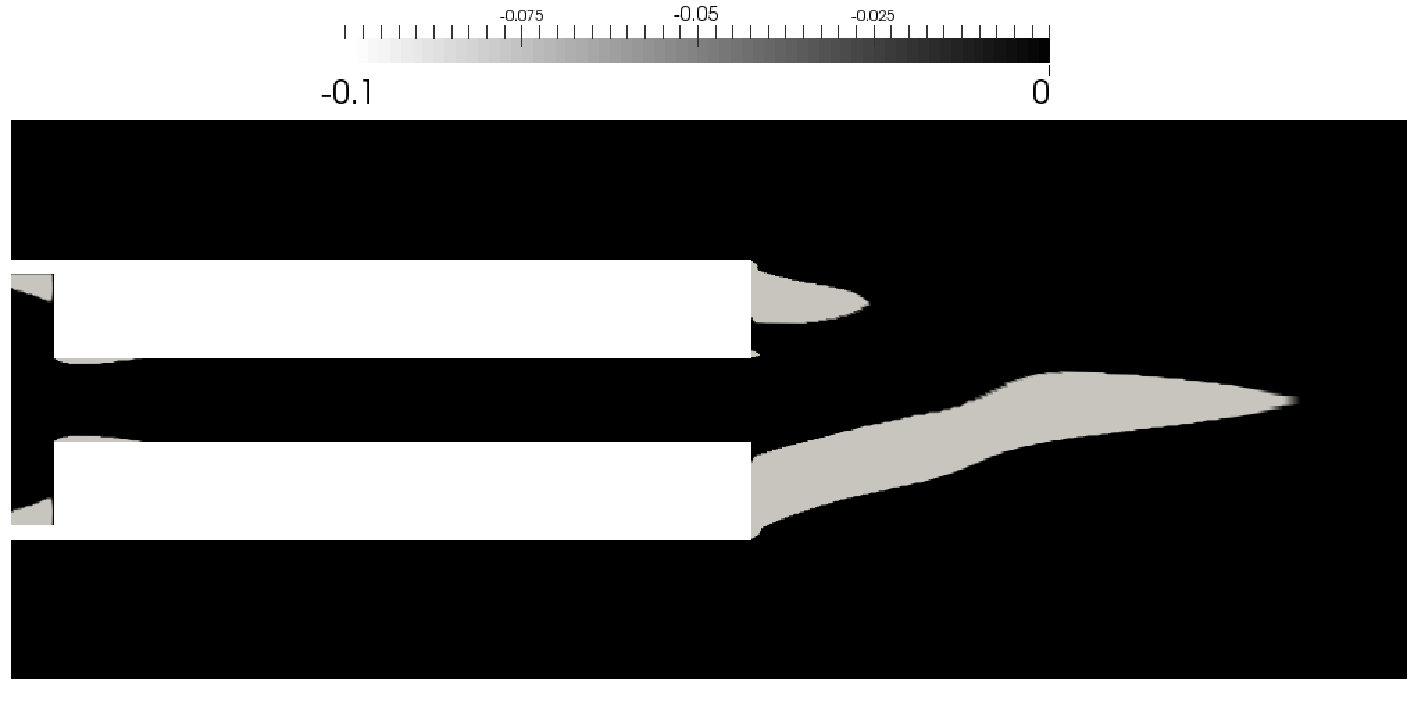}}
  \subfigure[22 $\%$ blowing rate]{\includegraphics{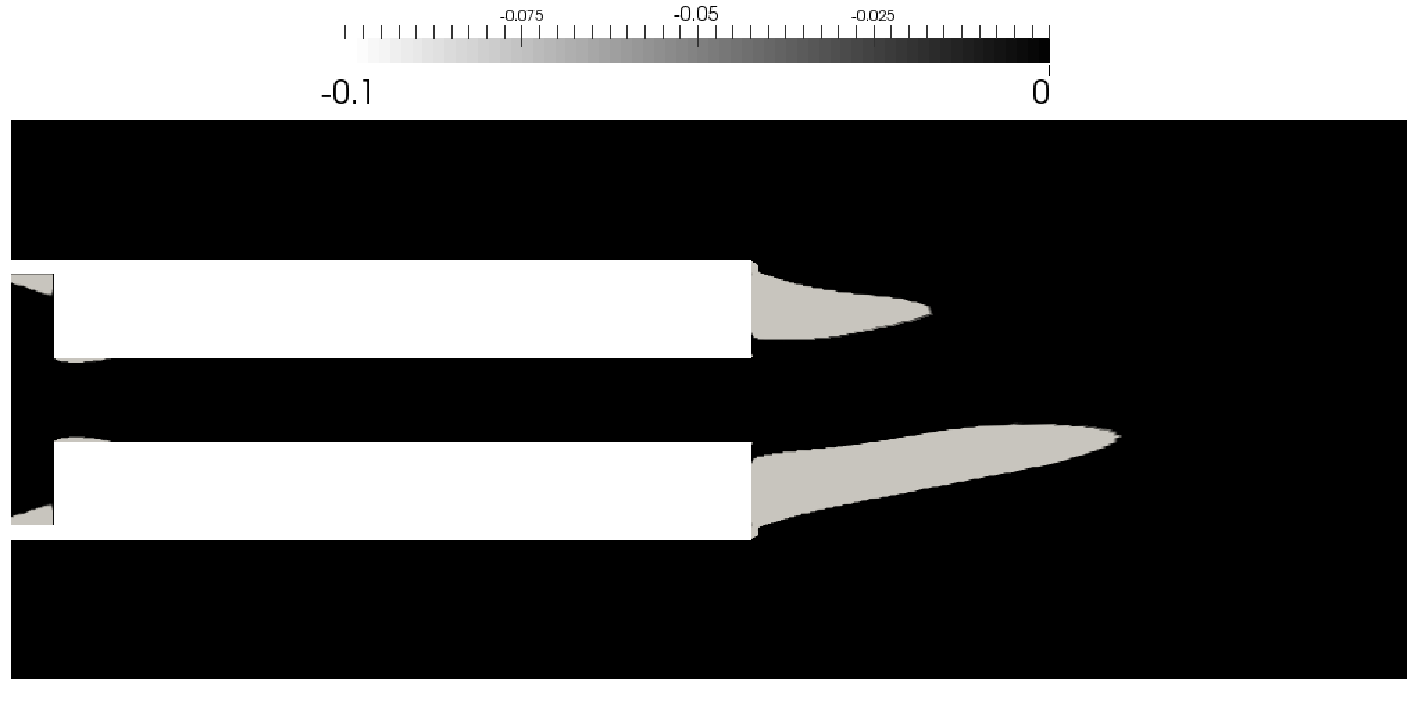}}
  \subfigure[33 $\%$ blowing rate]{\includegraphics{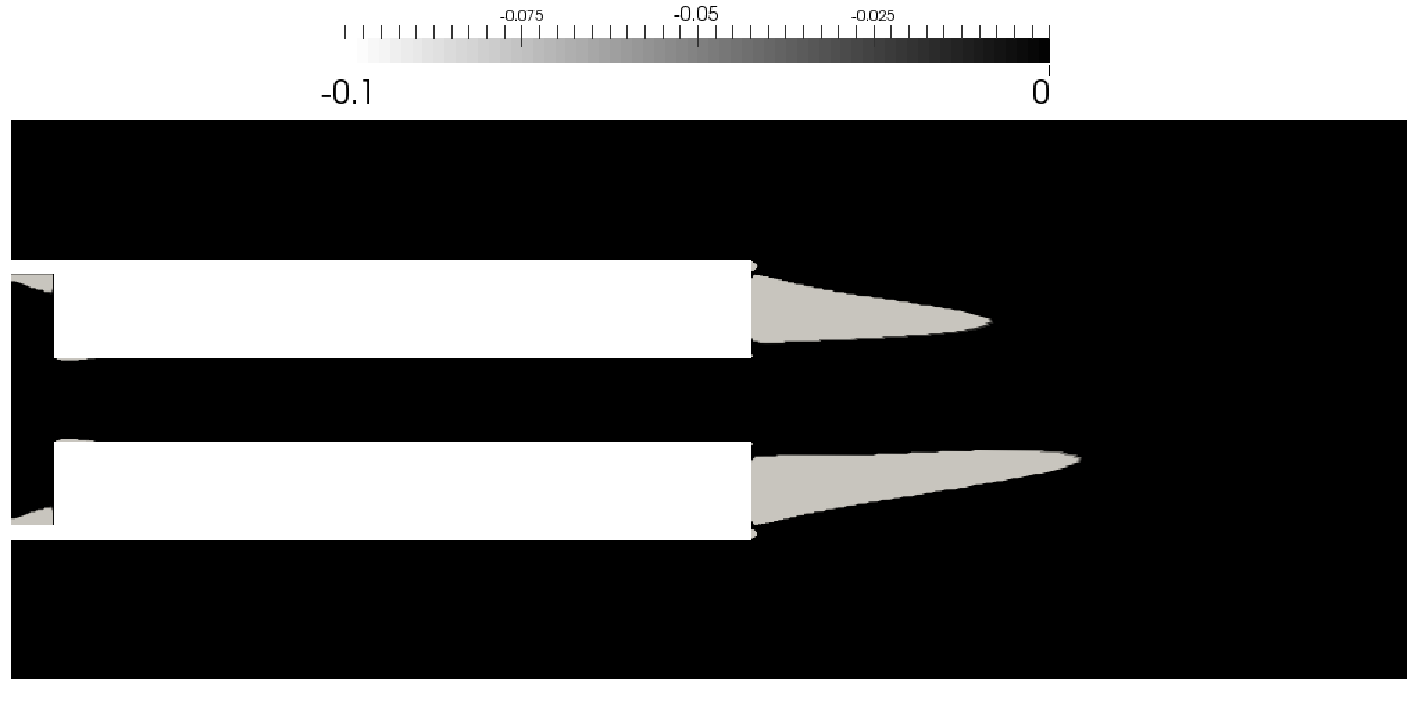}}
  \subfigure[42 $\%$ blowing rate]{\includegraphics{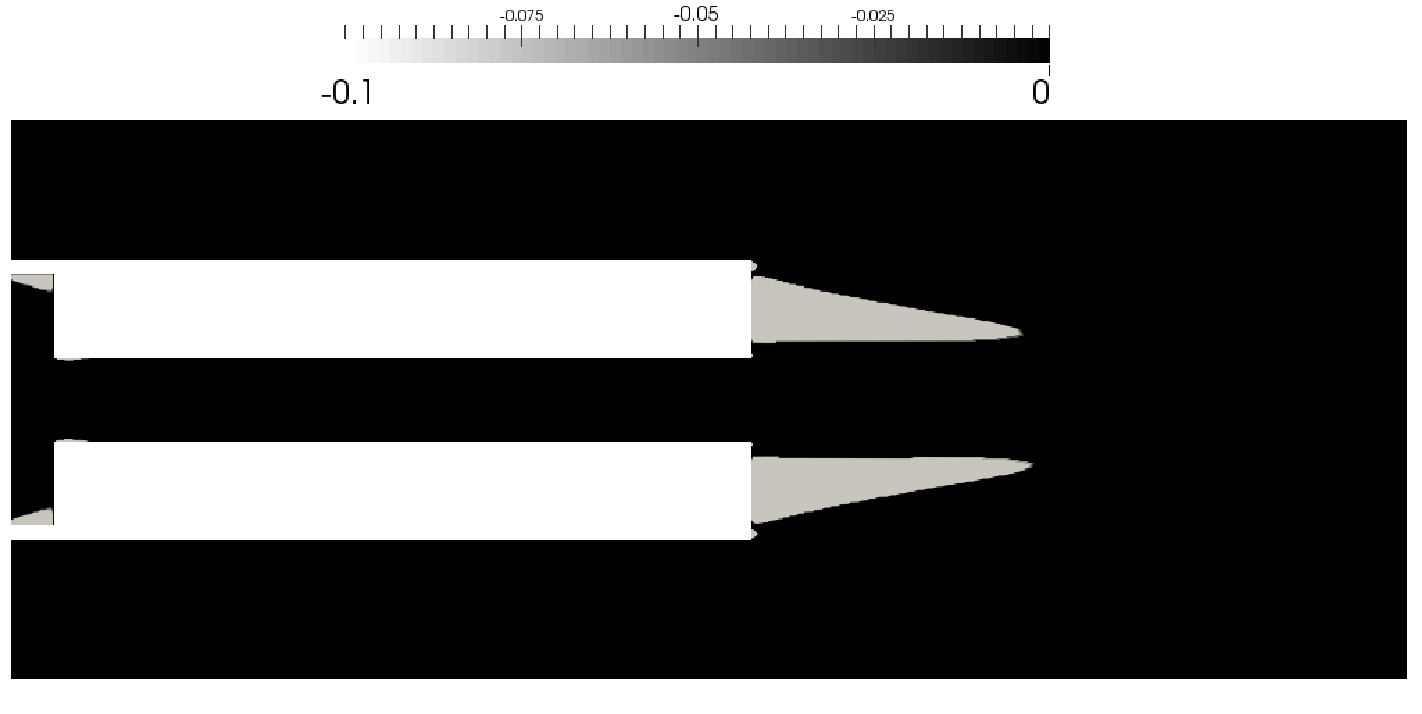}}
 \end{subfigmatrix}
 \caption{Dead air regions for different test cases and no bleeding configuration. Only negative velocities values are plotted.}
 \label{f:baseflow-deadair}
\end{figure}

\begin{figure}
\patchcmd{\subfigmatrix}{\hfill}{\hspace{0.8cm}}{}{}
 \begin{subfigmatrix}{2}
  \subfigure[18 $\%$ blowing rate]{\includegraphics[height=5cm]{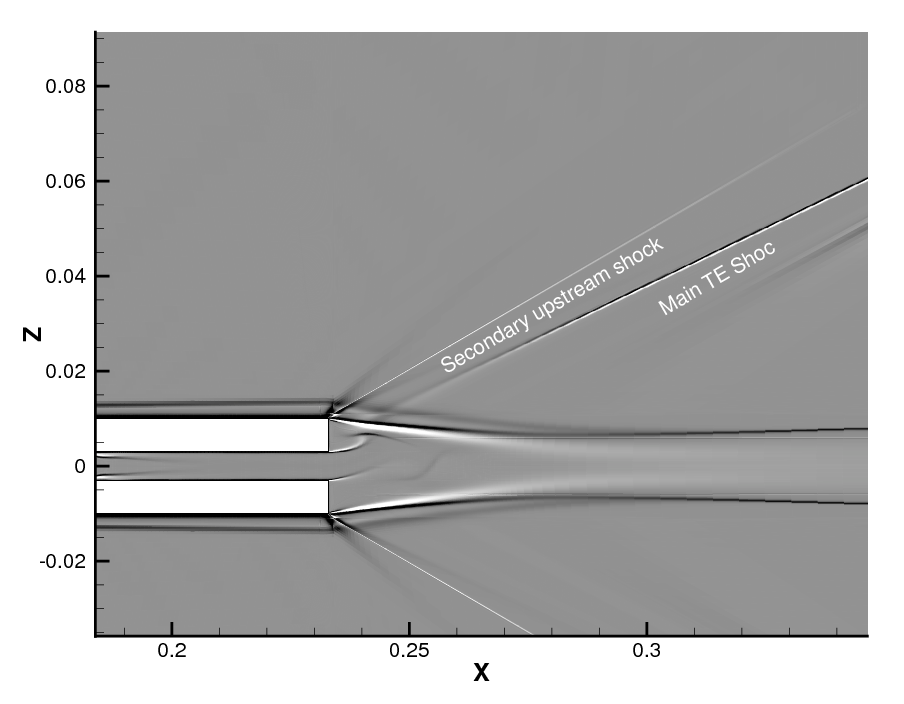}}
  \subfigure[45 $\%$ blowing rate]{\includegraphics[height=5cm]{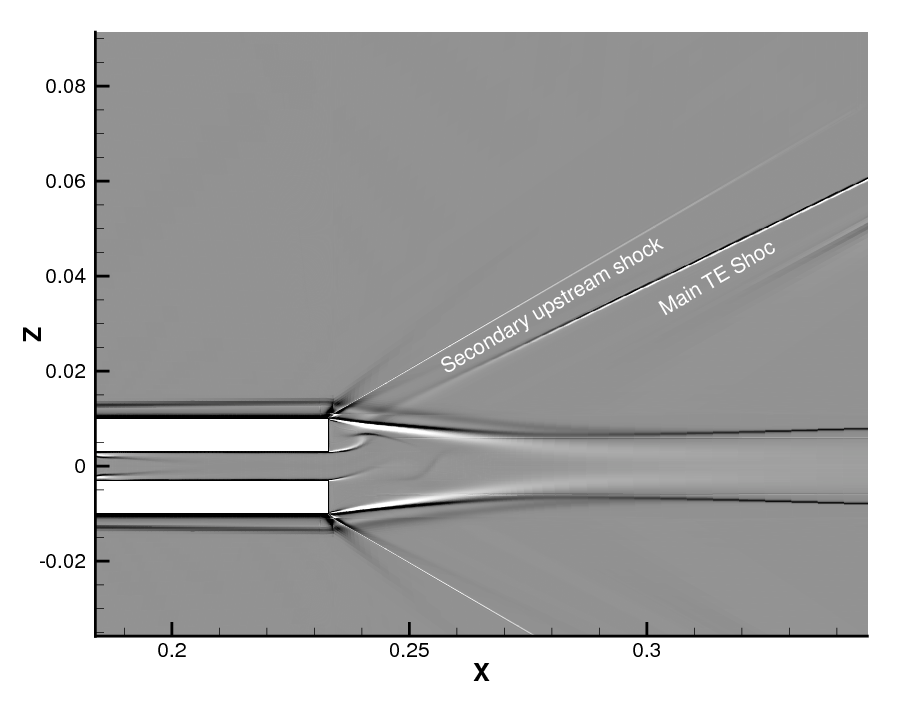}}
 \end{subfigmatrix}
 \caption{Shadowgraph contours for non-symmetric (a) and symmetric (b) flow configurations.}
 \label{f:baseflow-shadow}
\end{figure}

The analysis of the solutions obtained with the full-mesh allowed to understand the mechanisms of the non-symmetry. A detailed evolution of the recirculation areas is shown on Figure \ref{f:baseflow-deadair}, where only negatives velocities are plotted. As the jet is applied at the base region, the recirculation area is pushed downstream, splitting the recirculation region in three main areas, two at the trailing edge tips and a third one downstream. However, for blowing ratios above the bifurcation point one of the recirculation bubbles of the trailing edge joins the recirculation area downstream, increasing the pressure on that side and forcing the flow to change its direction in a \textit{Coanda effect} style. This change alters the wake structure, affecting the shocks angles and intensity. Larger blowing rates, however, stabilize the flow pushing the dead air zones towards the symmetry plane as more flow is entrained into the main jet, until the symmetric state is achieved again. The non-symmetric effect can also be appreciated from shadow-graph contour fields shown in Figure \ref{f:baseflow-shadow}, where it is easier to identify the weakening of the trailing edge shock wave and the appearance of a secondary upstream shock only in one side of the domain. The secondary upstream shock has its foot near the region where the purging flow contacts the trailing edge shear layers, so if the flow changes its direction and deflects to the contrary side, the secondary shock would swap side as well.

\subsection{Stability analysis} \label{sec:stab-analysis}

On Saracoglu et al.\cite{Saracoglu2013}, the pressure measurements were done over a URANS simulation, showing an asymmetrical state that was constant in time, suggesting that the difference on pressure had not a periodic oscillation behavior but a steady one. As shown on the flow topology analysis (Sec \ref{sec:flow-topo}), the cooling flow experiments a contraction at the plenum area, followed by a sudden expansion at the base region, limited on the upper and lower sides by the shear layers. According to the type of perturbation observed on the base flow (Fig. \ref{f:baseregiondata}-(a)), the geometry configuration and the results of Saracoglu\cite{Saracoglu2013}, it would be expected to find the physical eigenvalues on the imaginary axis, having a null pulsation. This kind of instability, known as a pitchfork bifurcation, would led to a steady asymmetric behavior of the recirculation regions at the trailing edge. Therefore, the focus was set on the region of the spectrum close to the origin and a zero value for the shift parameter was used on the preconditioning of the Jacobian matrix.

As previously explained, the base flows were computed using the half-domain mesh whereas for the stability analysis the full-domain mesh was used. All the simulations were performed using M4 mesh of the convergence study and the reduced domain DR3 for the stability analysis. Density, momentum, energy and turbulent viscosity of the perturbation fields were obtained from this analysis.

\begin{wrapfigure}[18]{r}{0.4\textwidth}
  \centering
    \includegraphics[width=0.4\textwidth]{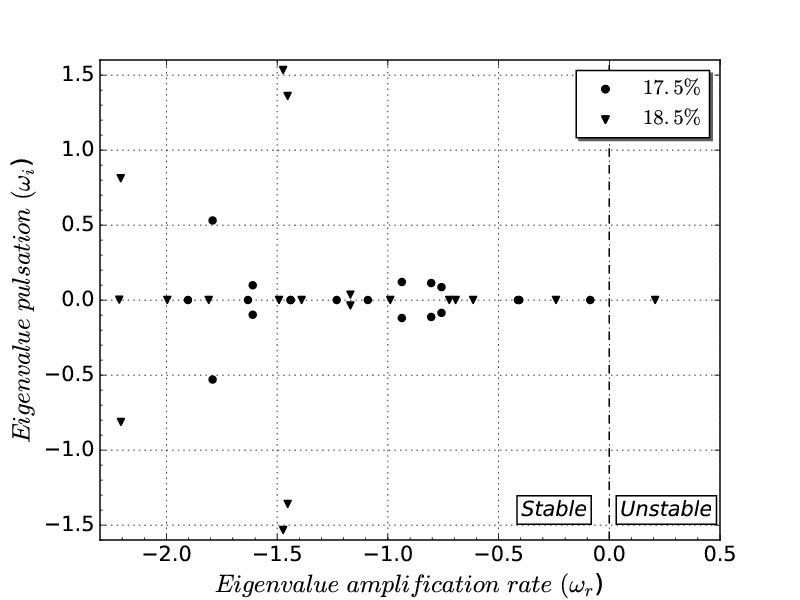}
  \caption{Evolution of the eigenvalue spectrum for a stable (17.5\%) and an unstable (18.5\%) configuration.}
  \label{f:spectros}
\end{wrapfigure}

\begin{figure}
\patchcmd{\subfigmatrix}{\hfill}{\hspace{0.8cm}}{}{}
	\centering
    \begin{subfigmatrix}{2}
        \subfigure[Eigenvalue evolution with the blowing rate]{\includegraphics[height=5cm,keepaspectratio]{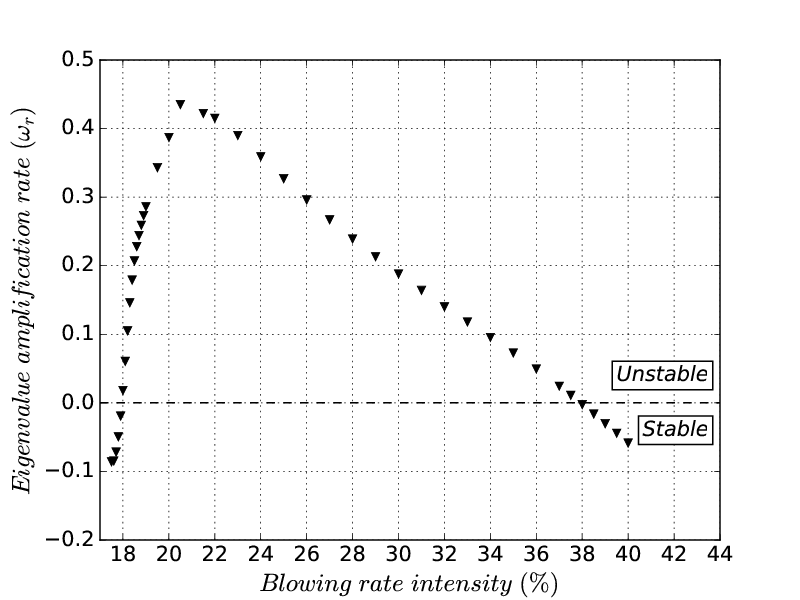}}
        \subfigure[Flow solution for a non-symmetric configuration]{\includegraphics[height=5cm,keepaspectratio]{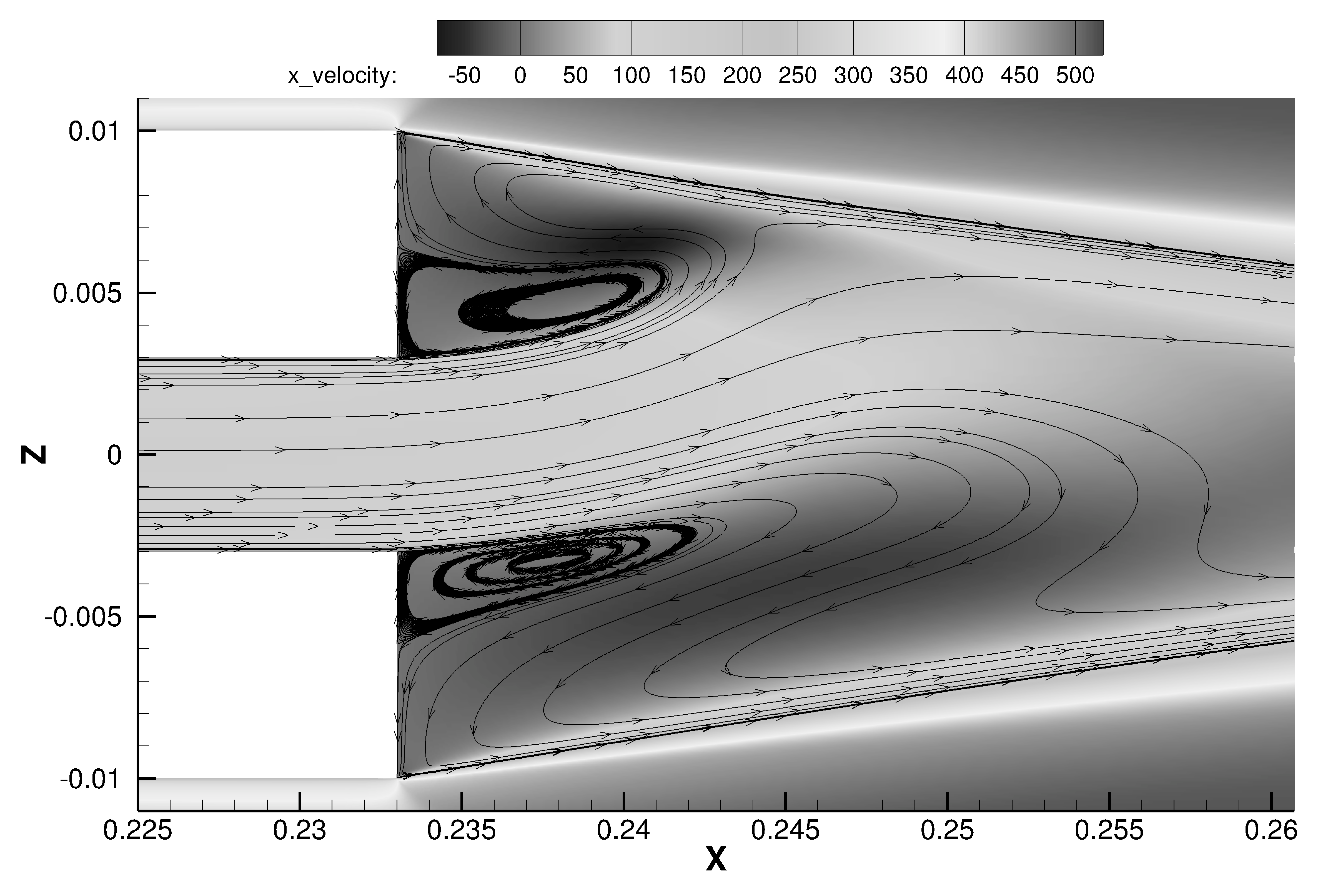}}
    \end{subfigmatrix}
    \caption{Relation between the global mode and the flow topology.}
 \label{f:stab-eigenval}
\end{figure}

For blowing rates below 18\%, all the eigenvalues of the spectrum remain on the stable region. When the blowing rate is increased, however, a single anti-symmetrical mode becomes unstable and its amplification rate evolves as a function of this parameter (Fig \ref{f:spectros}). This mode is directly related to the pressure bifurcation, its associate eigenvalue crossing the real axis when the purge intensity is above 18.1\%, and becoming stable again for a blowing rate of 38\% (Fig. \ref{f:stab-eigenval}). When the eigenvalue evolution is related to the base flow results, it can be observed that as the base region experiments changes on its flow topology, enlarging the recirculation bubbles of the trailing edge and showing an increment of the pressure drop at the wake (as shown in Figures \ref{f:baseflow-pressure} and \ref{f:baseflow-deadair}), the amplification rate of the unstable eigenvalue reaches its maximum value. After this point, the eigenvalue starts its damping until a blowing rate value of 38\%, where it becomes stable again at the same time the base flow recovers its symmetric state.

\begin{figure}
\patchcmd{\subfigmatrix}{\hfill}{\hspace{0.8cm}}{}{}
    \begin{subfigmatrix}{2}
         \subfigure[18 $\%$ blowing rate]{\includegraphics[width=0.45\textwidth]{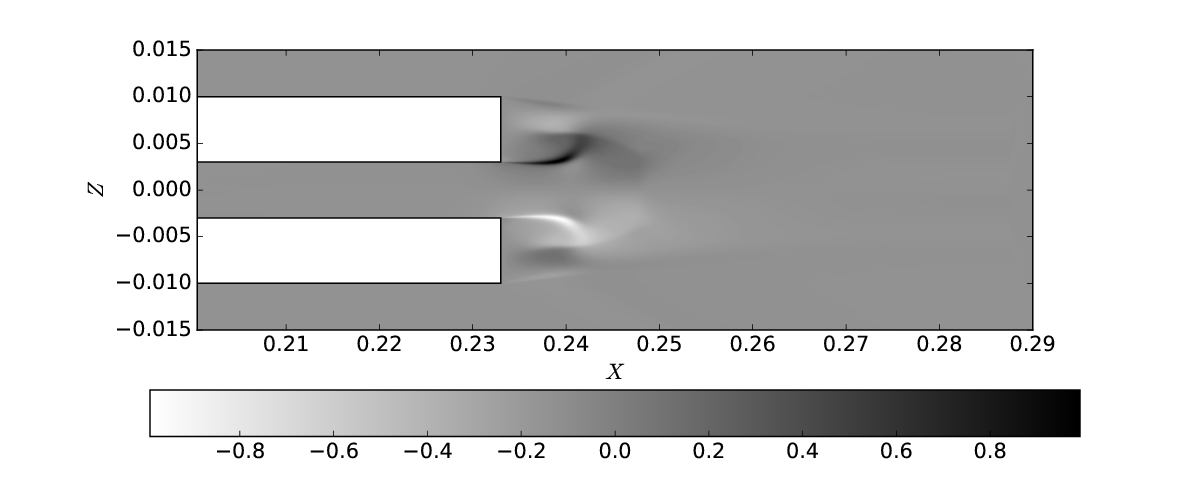}}
         \subfigure[20 $\%$ blowing rate]{\includegraphics[width=0.45\textwidth]{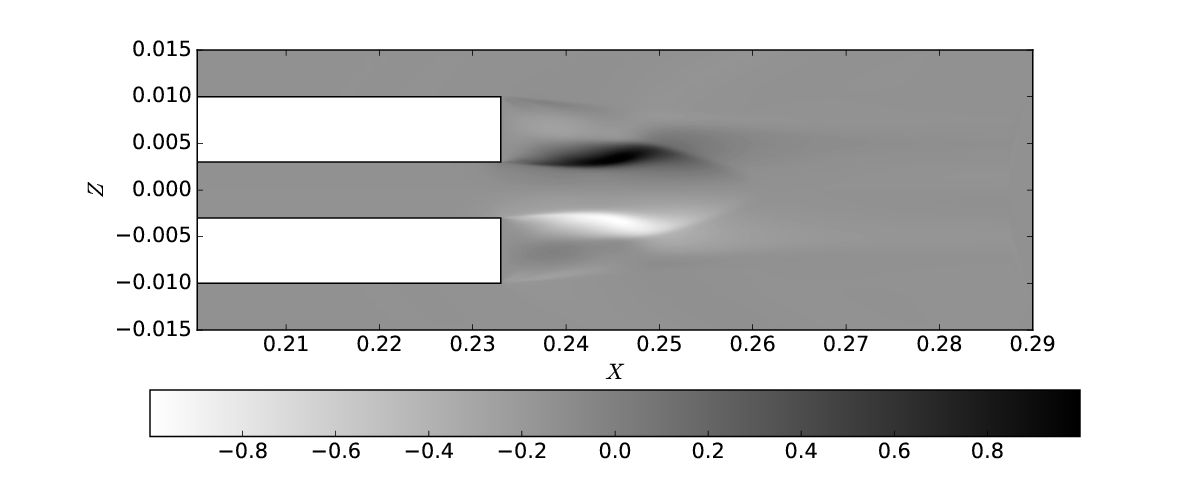}}
         \subfigure[30 $\%$ blowing rate]{\includegraphics[width=0.45\textwidth]{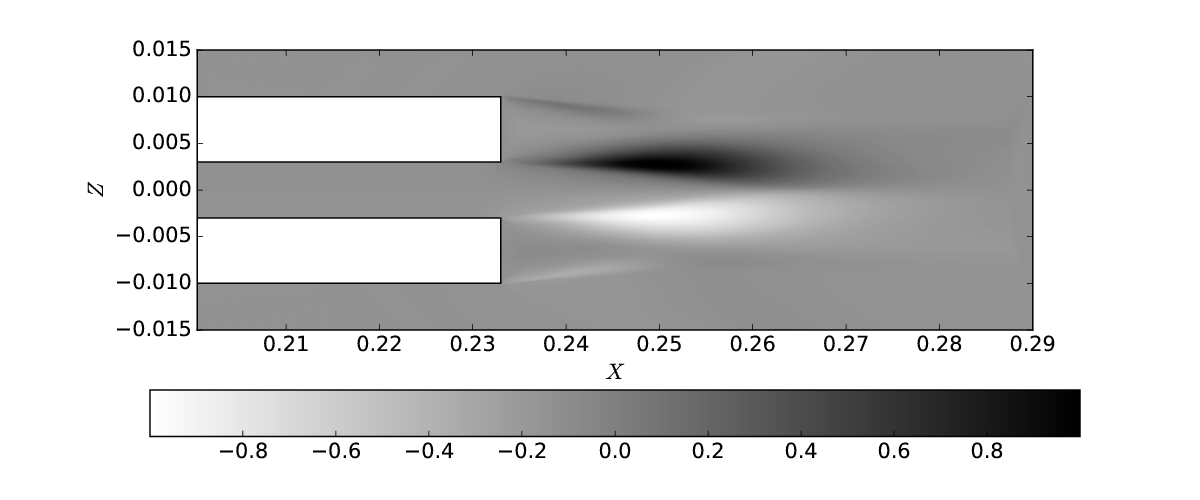}}
         \subfigure[40 $\%$ blowing rate]{\includegraphics[width=0.45\textwidth]{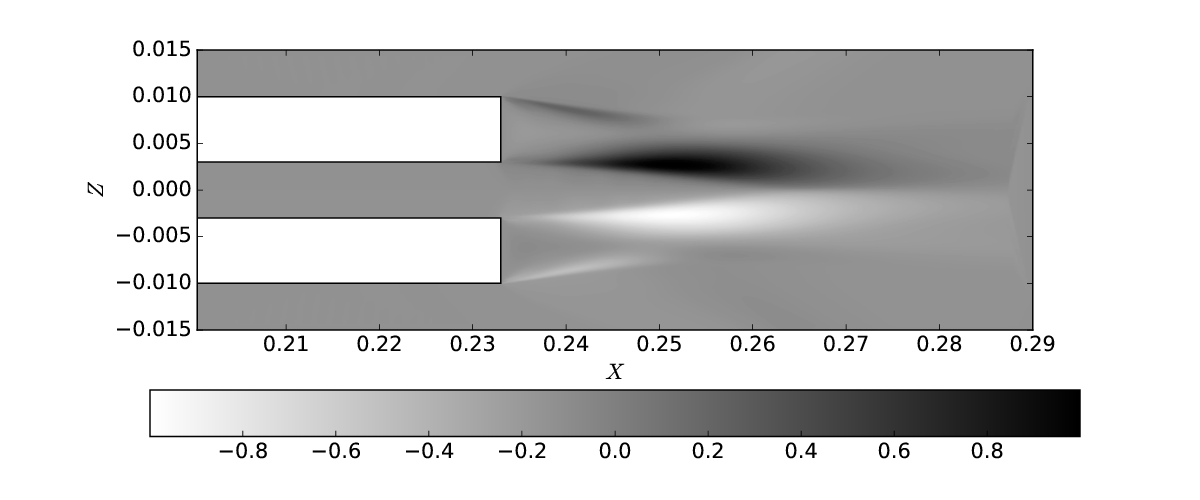}}
    \end{subfigmatrix}
    \caption{Evolution of the streamwise component of the eigenmode with the blowing rate.}
    \label{f:stab-eigenmodes}
\end{figure}

The eigenmode structure consists in two symmetrical lobe-shaped regions with anti-symmetrical streamwise perturbation components, and it changes its structure as long as the blowing rate increased, in consonance with the flow topology of the base flow (Fig. \ref{f:stab-eigenmodes}). At lower purge intensity cases, the mode appears concentrated at the exit of the purge channel and associated to the weak shear layers that occurred on the mixing of the base region and the purge flow. Subsequently, for larger blowing rates the mode changes its shape becoming more similar to the structures observed in common channel expansions\cite{Fani2012,Mizushima2001}, where it adopts the shape of two enlarged lobes bounded by the strong shear layers of the upper and lower tips of the trailing edge .

The behavior of the instability amplification rate is different at low and high blowing intensities. For a weak trailing edge blowing the rate of change of the amplification rate of the mode is high, and small variations of blowing rates would produce a rapid change of the global mode behavior from stable to unstable; at higher blowing rates the tendency is smoother and the changes in stability behavior are less abrupt. This fact can also explain the small differences between the RANS results and the stability analysis. Namely, according to RANS computations, the flow remains non-symmetric until a blowing rate of around 42\%, compared to a value of 38\% for the stability analysis. However, it is normally observed that, when the amplification rate of the instability is not large enough, the flow can remain in a ``unstable" situation for a long time, unless a perturbation is introduced in the flow triggering the new flow configuration\cite{Valero2014}. 

\section{Conclusions}

The phenomenon of the sudden expansion related instability is replicated in this context for the supersonic blowing trailing edge problem, a simplified version of the flow configuration close to the trailing edge of a turbomachinery turbine blade. The presence of a bifurcation from symmetric to non-symmetric flow configuration as a function of the jet blowing rate had been observed but not explained. Previous analysis of this geometrical model suggested that the bifurcation was generated by a sudden expansion mechanism\cite{Saracoglu2013}. This fact was confirmed on this work with the use of Global Stability Theory. 

The presented results link the non-symmetrical configuration with an unstable anti-symmetric eigenmode for a range of blowing rates, matching the start and end points of the pressure bifurcation with small error when compared with RANS results. Comparing the free stream to purge density ratio ($\rho_{purge} / \rho_f$) of RANS solutions and stability analysis, an error value of 0.5\% was found for low blowing rates, with an error of about 5\% for higher blowing rates. The eigenmode flow structures appeared to extend downstream in space as the purge intensity increased, revealing itself as a clear sudden expansion mode for higher blowing rates (having the same structure as those seen in Fani et al.\cite{Fani2012}), but more concentrated at the end of the cooling channel for lower blowing rates. It is on this range of blowing rates where the mode takes an structure more opened to the lower and upper sides, in a shear layer shape, and where it is more sensitive to changes in the blowing rate. 

When compared with a global mode related with a geometrical flow sudden expansion, its similarity in structure seemed clear despite the absence of viscous walls on the upper and lower sides aft the expansion that would bound the blow. In the configuration analyzed here, however, the flow was confined due the the strong pressure gradients generated by the shear layers present on the trailing edge that limit the base region. This shear layers changed its shape with the blowing rate intensity, modifying this way the boundaries of the sudden expansion and thus affecting the eigenmode structure.

As observed, base pressure magnitude inversely correlated with trailing edge shock intensity. The variation of the cooling flow purge initially resulted in an increment of the base pressure and thus a decrease on the shock strength; for  blowing rates higher than 25\% this tendency reversed, the base pressure decayed resulting in stronger shock waves (Fig. \ref{f:baseflow-pressure}). Due to the destabilizing effect of the pressure bifurcation, the region of maximum base pressure was affected by a variation on the pressure values of upper and lower sides at the trailing edge, which could furthermore abruptly change from one branch of the bifurcation to the contrary, generating strong loads on the turbine blades because of the change of the``polarity" of the pressure difference. Despite the apparently low difference between upper and lower pressure sides (only about a 3\%, as shown in Fig. \ref{f:baseregiondata}), the pressure bifurcation could had a strong effect on the surrounding flow topology and adjacent blades of the turbine. For blowing rates comprised between 18 and 38\%, the global instability forced the flow to deflect towards one of the shear layers (Fig. \ref{f:stab-eigenval}-(b)) generating a secondary shock wave upstream of the trailing edge shock in only one side (Fig. \ref{f:baseflow-shadow}). This additional shock system, placed in only one side of the blade, would change to the opposite side when the deflected purge flow changed its direction due to a bifurcation branch change, therefore generating additional and non-symmetrical loads on the turbine cascade. If this intermittent non symmetric shock system is replicated all over the turbine blades, considering independent changes and interactions with upstream and downstream stages of the turbine, the consequences in the aerodynamic loads can be very important. In order to avoid these risks, one should increase the blowing rate up to an state of no variation (symmetry), implying an important increment on base pressure losses and intensity of the trailing edge shock wave system. The non time dependent nature of the instability would made it difficult to be controlled by frequency modulated systems (as those proposed by Saracoglu et al.\cite{Saracoglu2012}), but may be suitable for passive control or upstream flow modulation. The possibility of controlling the pressure bifurcation avoiding the mentioned problems would allow to exploit lower blowing rates for base region modulation, without undesired consequences.

Understanding the mechanisms of the non-symmetric flow configurations present on a supersonic trailing edge, using a simplified model as the one described in this work, is the first step on preventing unwanted loads on turbine blades by Global Stability Analysis. The methodology here explained can be applied without changes to more realistic cases, where the pressure gradient between both sides of the airfoil can play an important role on damping/exciting the sudden expansion mode. A closer approximation to the problem could allow to damp its sources (the unstable eigenmode) and even develop optimization techniques with direct feedback from the eigenvalue analysis.


\section*{Funding Sources}
This research was carried out under the project SSeMID, which has received funding from the European Union Horizon2020 research and innovation programme under the Marie Skłodowska-Curie grant agreement No 675008.
\section*{Acknowledgments}
Authors thankfully acknowledge the computer resources, technical expertise and assistance provided by the Centro de Computaci\'on y Visualizaci\'on de Madrid (CeSViMa).


\bibliography{R1_BIB-Inyector}

\begin{thebibliography}{45}
\newcommand{\enquote}[1]{``#1''}
\providecommand{\natexlab}[1]{#1}
\providecommand{\url}[1]{\texttt{#1}}
\providecommand{\urlprefix}{URL }
\expandafter\ifx\csname urlstyle\endcsname\relax
  \providecommand{\doi}[1]{doi:\discretionary{}{}{}#1}\else
  \providecommand{\doi}{doi:\discretionary{}{}{}\begingroup
  \urlstyle{rm}\Url}\fi

\bibitem[{Saracoglu et~al.(2013)Saracoglu, Paniagua, Sanchez, and
  Rambaud}]{Saracoglu2013}
Saracoglu, B.~H., Paniagua, G., Sanchez, J., and Rambaud, P., \enquote{{Effects
  of blunt trailing edge flow discharge in supersonic regime},} \emph{Computers
  and Fluids}, Vol.~88, 2013, pp. 200--209.
\newblock \doi{10.1016/j.compfluid.2013.09.013}.

\bibitem[{Nash(1963)}]{Nash1963}
Nash, J.~F., \enquote{{A Review of Research on Two-Dimensional Base Flow},}
  \emph{ARC R\&M}, , No. 3323, 1963, pp. 1--25.

\bibitem[{Nash(1967)}]{Nash1967}
Nash, J.~F., \enquote{{A Discussion of Two-Dimensional Turbulent Base Flows},}
  \emph{ARC R\&M}, , No. 3468, 1967, pp. 1--46.

\bibitem[{Nash et~al.(1966)Nash, Quincey, and Callinan}]{Nash1966}
Nash, J.~F., Quincey, V.~G., and Callinan, J., \enquote{{Experiments on
  Two-Dimensional Base Flow at Subsonic and Transonic Speeds},} \emph{ARC
  R\&M}, , No. 3427, 1966, pp. 1--34.

\bibitem[{Hama(1968)}]{Hama1968}
Hama, F.~R., \enquote{{Experimental Studies on the Lip Shock},} \emph{AIAA
  Journal}, Vol.~6, No.~2, 1968, pp. 212--219.
\newblock \doi{10.2514/3.4480}.

\bibitem[{Sieverding et~al.(1979)Sieverding, Stanislas, and
  Snoek}]{Sieverding1979}
Sieverding, C.~H., Stanislas, M., and Snoek, J., \enquote{{The Base Pressure
  Problem in Transonic Turbine Cascades},} \emph{Journal of Engineering for
  Power}, Vol. 102, No.~3, 1979, pp. 711--718.
\newblock \doi{10.1115/1.3230330}.

\bibitem[{Denton and Xu(1990)}]{Denton1990}
Denton, J.~D., and Xu, L., \enquote{{The Trailing Edge Loss of Transonic
  Turbine Blades},} \emph{Journal of Turbomachinery}, Vol. 112, No.~2, 1990,
  pp. 277--285.
\newblock \doi{10.1115/1.2927648}.

\bibitem[{Herrin and Dutton(1994)}]{Herrin1994}
Herrin, J.~L., and Dutton, J.~C., \enquote{{Supersonic base flow experiments in
  the near wake of a cylindrical afterbody},} \emph{AIAA Journal}, Vol.~32,
  No.~1, 1994, pp. 77--83.
\newblock \doi{10.2514/3.11953}.

\bibitem[{Sandberg(2012)}]{Sandberg2012}
Sandberg, R.~D., \enquote{{Numerical investigation of turbulent supersonic
  axisymmetric wakes},} \emph{Journal of Fluid Mechanics}, Vol. 702, 2012, pp.
  488--520.
\newblock \doi{10.1017/jfm.2012.201}.

\bibitem[{Sandberg and Fasel(2006)}]{Sandberg2006}
Sandberg, R.~D., and Fasel, H.~F., \enquote{{Numerical investigation of
  transitional supersonic axisymmetric wakes},} \emph{Journal of Fluid
  Mechanics}, Vol. 563, 2006, pp. 1--41.
\newblock \doi{10.1017/S0022112006000899}.

\bibitem[{Saracoglu et~al.(2012)Saracoglu, Paniagua, Salvadori, Tomasoni, Duni,
  Yasa, and Miranda}]{Saracoglu2012}
Saracoglu, B.~H., Paniagua, G., Salvadori, S., Tomasoni, F., Duni, S., Yasa,
  T., and Miranda, A., \enquote{{Trailing edge shock modulation by pulsating
  coolant ejection},} \emph{Applied Thermal Engineering}, Vol.~48, 2012, pp.
  1--10.
\newblock \doi{10.1016/j.applthermaleng.2012.04.036}.

\bibitem[{Raffel and Kost(1998)}]{Raffel1998}
Raffel, M., and Kost, F., \enquote{{Investigation of aerodynamic effects of
  coolant ejection at the trailing edge of a turbine blade model by PIV and
  pressure measurements},} \emph{Experiments in Fluids}, Vol.~24, No. 5-6,
  1998, pp. 447--461.
\newblock \doi{10.1007/s003480050194}.

\bibitem[{Kost and Holmes(1985)}]{Kost1985}
Kost, F.~H., and Holmes, A.~T., \enquote{Aerodynamic effect of coolant ejection
  in the rear part of transonic rotor blades,} \emph{AGARD Heat transfer and
  Cooling in Gas Turbines 12 p (SEE N86-29823 21-07)}, 1985.

\bibitem[{Bohn et~al.(1995)Bohn, Becker, Behnke, and Bonhoff}]{Bohn1995}
Bohn, D.~E., Becker, V.~J., Behnke, K.~D., and Bonhoff, B.~F.,
  \enquote{{Experimental and numerical investigations of the aerodynamical
  effects of coolant injection through the trailing edge of a guide vane},}
  \emph{American Society of Mechanical Engineers (Paper), in Proc. of ASME
  International Gas Turbine and Aeroengine Congress and Exposition}, Vol.
  95-GT-26, Houston, Texas, 1995.
\newblock \doi{10.1115/95-GT-026}.

\bibitem[{Wang and Zhao(2013)}]{Wang2013}
Wang, Y., and Zhao, L., \enquote{{Investigation on the effect of trailing edge
  ejection on a turbine cascade},} \emph{Applied Mathematical Modelling},
  Vol.~37, No.~9, 2013, pp. 6254--6265.
\newblock \doi{10.1016/j.apm.2013.01.023}.

\bibitem[{Iorio et~al.(2014)Iorio, Gonz{\'{a}}lez, and Ferrer}]{Iorio2014}
Iorio, M.~C., Gonz{\'{a}}lez, L.~M., and Ferrer, E., \enquote{{Direct and
  adjoint global stability analysis of turbulent transonic flows over a
  NACA0012 profile},} \emph{International Journal for Numerical Methods in
  Fluids}, Vol.~76, No.~3, 2014, pp. 147--168.
\newblock \doi{10.1002/fld.3929}.

\bibitem[{Iorio et~al.(2015)Iorio, Gonz{\'{a}}lez, and
  Martinez-Cava}]{Iorio2015}
Iorio, M.~C., Gonz{\'{a}}lez, L.~M., and Martinez-Cava, A., \enquote{{Global
  Stability Analysis of a Compressible Turbulent Flow around a High-Lift
  Configuration},} \emph{AIAA Journal}, Vol.~54, No.~2, 2015, pp. 373--385.
\newblock \doi{10.2514/1.J054211}.

\bibitem[{Sartor et~al.(2015)Sartor, Mettot, Bur, and Sipp}]{Sartor2015}
Sartor, F., Mettot, C., Bur, R., and Sipp, D., \enquote{{Unsteadiness in
  transonic shock-wave/boundary-layer interactions: experimental investigation
  and global stability analysis},} \emph{Journal of Fluid Mechanics}, Vol. 781,
  2015, pp. 550--577.
\newblock \doi{10.1017/jfm.2015.510}.

\bibitem[{Barkley(2006)}]{Barkley2006}
Barkley, D., \enquote{{Linear analysis of the cylinder wake mean flow},}
  \emph{Europhysics Letters (EPL)}, Vol.~75, No.~5, 2006, pp. 750--756.
\newblock \doi{10.1209/epl/i2006-10168-7}.

\bibitem[{Oberleithner et~al.(2014)Oberleithner, Rukes, and
  Soria}]{Oberleithner2014}
Oberleithner, K., Rukes, L., and Soria, J., \enquote{{Mean flow stability
  analysis of oscillating jet experiments},} \emph{Journal of Fluid Mechanics},
  Vol. 757, 2014.
\newblock \doi{10.1017/jfm.2014.472}.

\bibitem[{Sipp and Lebedev(2007)}]{Sipp2007}
Sipp, D., and Lebedev, A., \enquote{{Global stability of base and mean flows: a
  general approach and its applications to cylinder and open cavity flows},}
  \emph{Journal of Fluid Mechanics}, Vol. 593, 2007, pp. 333--358.
\newblock \doi{10.1017/S0022112007008907}.

\bibitem[{Grilli et~al.(2012)Grilli, Schmid, Hickel, and Adams}]{Grilli2012}
Grilli, M., Schmid, P.~J., Hickel, S., and Adams, N.~A., \enquote{{Analysis of
  unsteady behaviour in shockwave turbulent boundary layer interaction},}
  \emph{Journal of Fluid Mechanics J. Fluid Mech}, Vol. 700, No. 700, 2012, pp.
  16--28.
\newblock \doi{10.1017/jfm.2012.37}.

\bibitem[{Lashgari et~al.(2014)Lashgari, Tammisola, Citro, Juniper, and
  Brandt}]{Lashgari2014}
Lashgari, I., Tammisola, O., Citro, V., Juniper, M.~P., and Brandt, L.,
  \enquote{{The planar X-junction flow: stability analysis and control},}
  \emph{Journal of Fluid Mechanics}, Vol. 753, No. August, 2014, pp. 1--28.
\newblock \doi{10.1017/jfm.2014.364}.

\bibitem[{Wilcox(2008)}]{Wilcox2008}
Wilcox, D.~C., \enquote{{Formulation of the k-w Turbulence Model Revisited},}
  \emph{AIAA Journal}, Vol.~46, No.~11, 2008, pp. 2823--2838.
\newblock \doi{10.2514/1.36541}.

\bibitem[{Schwamborn et~al.(2006)Schwamborn, Gerhold, and Heinrich}]{taucode}
Schwamborn, D., Gerhold, T., and Heinrich, R., \enquote{{The DLR Tau-code:
  Recent Applications in Research and Industry},} \emph{Proceedings of the
  European Conference on Computational Fluid Dynamics, ECOMAS 2006}, , No. June
  2017, 2006.

\bibitem[{Liou and Steffen(1993)}]{liou1993}
Liou, M.-S., and Steffen, C.~J., \enquote{A new flux splitting scheme,}
  \emph{Journal of Computational physics}, Vol. 107, No.~1, 1993, pp. 23--39.
\newblock \doi{10.1006/jcph.1993.1122}.

\bibitem[{Anderson and Wendt(1995)}]{Anderson1995}
Anderson, J.~D., and Wendt, J., \emph{{Computational fluid dynamics}}, Vol.
  206, Springer, 1995.

\bibitem[{Chomaz(2005)}]{Chomaz2005}
Chomaz, J.-m., \enquote{{GLOBAL INSTABILITIES IN SPATIALLY DEVELOPING FLOWS:
  Non-Normality and Nonlinearity},} \emph{Annual Review of Fluid Mechanics},
  Vol.~37, No.~1, 2005, pp. 357--392.
\newblock \doi{10.1146/annurev.fluid.37.061903.175810}.

\bibitem[{Theofilis(2011)}]{Theofilis2011}
Theofilis, V., \enquote{{Global Linear Instability},} \emph{Annual Review of
  Fluid Mechanics}, Vol.~43, No.~1, 2011, pp. 319--352.
\newblock \doi{10.1146/annurev-fluid-122109-160705}.

\bibitem[{Luchini and Bottaro(2014)}]{Luchini2014}
Luchini, P., and Bottaro, A., \enquote{{Adjoint Equations in Stability
  Analysis},} \emph{Annual Review of Fluid Mechanics}, Vol.~46, No.~1, 2014,
  pp. 493--517.
\newblock \doi{10.1146/annurev-fluid-010313-141253}.

\bibitem[{Taira et~al.(2017)Taira, Brunton, Dawson, Rowley, Colonius, McKeon,
  Schmidt, Gordeyev, Theofilis, and Ukeiley}]{Taira2017}
Taira, K., Brunton, S.~L., Dawson, S. T.~M., Rowley, C.~W., Colonius, T.,
  McKeon, B.~J., Schmidt, O.~T., Gordeyev, S., Theofilis, V., and Ukeiley,
  L.~S., \enquote{{Modal Analysis of Fluid Flows: An Overview},} \emph{AIAA
  Journal}, Vol.~55, No.~12, 2017, pp. 4013--4041.
\newblock \doi{10.2514/1.J056060}.

\bibitem[{Meseguer-Garrido et~al.(2014)Meseguer-Garrido, de~Vicente, Valero,
  and Theofilis}]{Meseguer2014}
Meseguer-Garrido, F., de~Vicente, J., Valero, E., and Theofilis, V.,
  \enquote{{On linear instability mechanisms in incompressible open cavity
  flow},} \emph{Journal of Fluid Mechanics}, Vol. 752, No. 2014, 2014, pp.
  219--236.
\newblock \doi{10.1017/jfm.2014.253}.

\bibitem[{Gonz{\'{a}}lez et~al.(2011)Gonz{\'{a}}lez, Ahmed, K{\"{u}}hnen,
  Kuhlmann, and Theofilis}]{Gonzalez2011}
Gonz{\'{a}}lez, L.~M., Ahmed, M., K{\"{u}}hnen, J., Kuhlmann, H., and
  Theofilis, V., \enquote{{Three-dimensional flow instability in a lid-driven
  isosceles triangular cavity},} \emph{J. Fluid Mech}, Vol. 675, 2011, pp.
  369--396.
\newblock \doi{10.1017/S002211201100022X}.

\bibitem[{Gorbachova et~al.(2016)Gorbachova, Valero, Paniagua, Martinez-Cava,
  and Saracoglu}]{Gorbachova2016}
Gorbachova, Y., Valero, E., Paniagua, G., Martinez-Cava, A., and Saracoglu,
  B.~H., \enquote{Study of the attainable flow topologies in a supersonic blunt
  trailing edge at various blowing ratios,} \emph{54th AIAA Aerospace Sciences
  Meeting, AIAA SciTech Forum, (AIAA 2016-0909)}, American Institute of
  Aeronautics and Astronautics, 2016.
\newblock \doi{10.2514/6.2016-0909}.

\bibitem[{Fani et~al.(2012)Fani, Camarri, and Salvetti}]{Fani2012}
Fani, A., Camarri, S., and Salvetti, M.~V., \enquote{{Stability analysis and
  control of the flow in a symmetric channel with a sudden expansion},}
  \emph{Physics of Fluids}, Vol.~24, No.~8, 2012, p. 084102.
\newblock \doi{10.1063/1.4745190}.

\bibitem[{Fraysse et~al.(2013)Fraysse, Valero, and Rubio}]{Fraysse2013}
Fraysse, F., Valero, E., and Rubio, G., \enquote{{Quasi-a priori truncation
  error estimation and higher order extrapolation for non-linear partial
  differential equations},} \emph{Journal of Computational Physics}, Vol. 253,
  2013, pp. 389--404.
\newblock \doi{10.1016/j.jcp.2013.07.018}.

\bibitem[{Lehoucq et~al.(1997)Lehoucq, Sorensen, and Yang}]{ARPACK}
Lehoucq, R.~B., Sorensen, D.~C., and Yang, C., \enquote{ARPACK Users Guide:
  Solution of Large Scale Eigenvalue Problems by Implicitly Restarted Arnoldi
  Methods.} , 1997.

\bibitem[{Saad(2011)}]{Saad2011}
Saad, Y., \emph{{Numerical Methods for Large Eigenvalue Problems}}, Society for
  Industrial and Applied Mathematics, 2011.
\newblock \doi{10.1137/1.9781611970739}.

\bibitem[{{De Vicente} et~al.(2011){De Vicente}, Rodr{\'{i}}guez, Theofilis,
  and Valero}]{DeVicente2011}
{De Vicente}, J., Rodr{\'{i}}guez, D., Theofilis, V., and Valero, E.,
  \enquote{{Stability analysis in spanwise-periodic double-sided lid-driven
  cavity flows with complex cross-sectional profiles},} \emph{Computers and
  Fluids}, Vol.~43, No.~1, 2011, pp. 143--153.
\newblock \doi{10.1016/j.compfluid.2010.09.033}.

\bibitem[{Ferrer et~al.(2014)Ferrer, de~Vicente, and Valero}]{Ferrer2014}
Ferrer, E., de~Vicente, J., and Valero, E., \enquote{{Low cost 3D global
  instability analysis and flow sensitivity based on dynamic mode decomposition
  and high-order numerical tools},} \emph{International Journal for Numerical
  Methods in Fluids}, Vol.~76, No.~3, 2014, pp. 169--184.
\newblock \doi{10.1002/fld.3930}.

\bibitem[{Browne et~al.(2014)Browne, Rubio, Ferrer, and Valero}]{Browne2014}
Browne, O. M.~F., Rubio, G., Ferrer, E., and Valero, E., \enquote{{Sensitivity
  analysis to unsteady perturbations of complex flows: a discrete approach},}
  \emph{International Journal for Numerical Methods in Fluids}, Vol.~76,
  No.~12, 2014, pp. 1088--1110.
\newblock \doi{10.1002/fld.3962}.

\bibitem[{Amestoy et~al.(2001)Amestoy, Duff, L'Excellent, and Koster}]{MUMPS}
Amestoy, P.~R., Duff, I.~S., L'Excellent, J.-Y., and Koster, J., \enquote{A
  Fully Asynchronous Multifrontal Solver Using Distributed Dynamic Scheduling,}
  \emph{SIAM Journal on Matrix Analysis and Applications}, Vol.~23, No.~1,
  2001, pp. 15--41.
\newblock \doi{10.1137/S0895479899358194}.

\bibitem[{Sanvido et~al.(2018)Sanvido, Garicano-Mena, de~Vicente, and
  Valero}]{Sanvido2017}
Sanvido, S., Garicano-Mena, J., de~Vicente, J., and Valero, E.,
  \enquote{Critical Assessment of Domain Reduction Strategies for Global
  Stability Analisis,} \emph{Submitted for publication to Communications in
  Computational Physics}, 2018.

\bibitem[{Mizushima and Shiotani(2001)}]{Mizushima2001}
Mizushima, J., and Shiotani, Y., \enquote{{Transitions and instabilities of
  flow in a symmetric channel with a suddenly expanded and contracted part},}
  \emph{Journal of Fluid Mechanics}, Vol. 434, 2001, pp. 355--369.
\newblock \doi{10.1017/S0022112001003743}.

\bibitem[{Valero et~al.(2014)Valero, Ferrer, and de~Vicente}]{Valero2014}
Valero, E., Ferrer, E., and de~Vicente, J., \enquote{{Numerical Methods for
  Direct Numerical Simulation and Stability Analysis},} \emph{Progress in flow
  instability analysis and laminar-turbulent transition modeling, VKI LS
  2014-07}, edited by E.~Valero and F.~Pinna, Von Karman Institute for Fluid
  Dynamics, 2014.

\end{thebibliography}

\end{document}